\documentclass[11pt]{article}
\usepackage{verbatim,amsmath,amssymb}
\usepackage{epsfig,float,color}
\usepackage{epstopdf}
\usepackage{geometry}
\usepackage{setspace}
\usepackage{wrapfig}
\usepackage{hyperref}
\usepackage[utf8]{inputenc}
\usepackage{graphicx}
\usepackage[font={footnotesize}]{caption}
\usepackage[font={footnotesize}]{subcaption}
\usepackage{footnote}
\usepackage{multicol}
\usepackage{multirow}
\usepackage{enumitem}   
\usepackage{soul}
\usepackage{framed}
\usepackage{tikz,pgfplots}
\usepackage{bm}
\usepackage{siunitx}
\usepackage[makeroom]{cancel} 
\usepackage[numbers]{natbib}
\usepackage{hyperref}

\newenvironment{rcases}
{\left.\begin{aligned}}
	{\end{aligned}\right\rbrace}

\definecolor{darkblue}{rgb}{0,0,1}
\hypersetup{pdftex=true, colorlinks=true, breaklinks=true, linkcolor=darkblue, menucolor=darkblue, pagecolor=darkblue, citecolor=darkblue, urlcolor=darkblue}

\newcommand{\matr}[1]{\mathbf{#1}}
\newcommand{\mb}[1]{\mathbf{#1}} 
\newcommand{\pd}[2]{\frac{\partial #1}{\partial #2}}
\newcommand{\trr}[1]{{#1}^{\!\top}}

\newcommand{\inv}[1]{{#1}^{\text{-}1}}

\usetikzlibrary{calc, pgfplots.groupplots, backgrounds}
\usepgfplotslibrary{colorbrewer}
\usepackage{standalone}
\begin{document}
	
	\begin{center}
		\Large{\bf{On topology optimization of design-dependent pressure-loaded three-dimensional  structures and compliant mechanisms}}\\
		
	\end{center}
	
	\begin{center}

		\large{Prabhat Kumar$^{\ast,\,\ddagger,\,}$\footnote{Corresponding author: prabhatk@iisc.ac.in,\,prabhatkumar.rns@gmail.com} and 	Matthijs Langelaar$^\dagger$}
		\vspace{4mm}
		
		\small{\textit{$^\ast$Department of Mechanical Engineering, Indian Institute of Science,
				Bangalore 560012, Karnataka, India}}\\
		\small{\textit{$^\ddagger$Department of Mechanical Engineering, Solid Mechanics, Technical University of Denmark,
				2800 Kgs. Lyngby, Denmark}}
			
		\small{\textit{$^\dagger$Department of Precision and Microsystems Engineering, Delft University of Technology, 2628CD Delft, The Netherlands}}

			\vspace{4mm}
		Published\footnote{This pdf is the personal version of an article whose final publication is available at \href{https://onlinelibrary.wiley.com/doi/10.1002/nme.6618}{Int J Numer Methods Eng.}}\,\,\,in \textit{Int J Numer Methods Eng.}, 
		\href{https://onlinelibrary.wiley.com/doi/10.1002/nme.6618}{ https://doi.org/10.1002/nme.6618} \\
		Submitted on 04~September 2020, Revised on 17~November 2020, Accepted on 21~December 2020
		
	\end{center}
	
	\vspace{2mm}
	\rule{\linewidth}{.10mm}
	{\bf Abstract:}
	 This paper presents a density-based topology optimization method for designing  three-dimensional (3D) compliant mechanisms and loadbearing structures with design-dependent pressure loading. Instead of interface-tracking techniques, the Darcy law in conjunction with a drainage term is employed to obtain pressure field as a function of the design vector. To ensure continuous transition of pressure loads as the design evolves, the flow coefficient of a finite element is defined using a smooth Heaviside function. The obtained pressure field is converted into consistent nodal loads  using a transformation matrix. The presented approach employs the standard finite element formulation and also, allows consistent and computationally inexpensive calculation of load sensitivities using the adjoint-variable method. For compliant mechanism design, a multi-criteria objective is minimized, whereas minimization of compliance  is performed for designing loadbearing structures. Efficacy and robustness of the presented approach is demonstrated by designing various pressure-actuated 3D compliant mechanisms and structures.  \\
	

	\vspace{-4mm}
	\rule{\linewidth}{.10mm}

\section{Introduction}\label{Sec:Intro}

Nowadays, the use of topology optimization (TO) approaches in a wide variety of design problems for different applications involving single and/or multi-physics is continuously growing because of their proven capability and efficacy \cite{sigmund2013topology}. These methods determine an optimized material layout for a given design problem by extremizing the desired objective(s) under a given set of constraints. Based on the considered loading behavior, they can be classified into approaches involving \textit{design-independent} (invariant) loads and methods considering \textit{design-dependent} forces. The latter situation often arises in case of aerodynamic loads, hydrodynamic loads and/or hydrostatic pressure loads, in various applications including aircraft, pumps, ships and pneumatically actuated soft robotics \cite{kenway2014multipoint,rus2015design,yang2016overview}. Many TO  methods exist for the former loading scenarios, whereas only few methods considering design-dependent loading behaviors have been reported in TO \cite{kumar2020topology}. Design-dependent loads alter their location, direction and/or magnitude as optimization progresses and thus, pose unique challenges~\cite{Hammer2000}. Those challenges get even more pronounced in a 3D TO setting \cite{du2004topologicalb,zhang2010topology}.  Here, our motive is to present an efficient and robust TO method suitable for 3D design problems including loadbearing structures and small deformation compliant mechanisms involving design-dependent pressure loads.

Compliant mechanisms (CMs), monolithic structures incorporating flexible regions,
rely on their elastic deformation to achieve their mechanical tasks in response to  external stimuli. These mechanisms furnish many advantages over their rigid-body counterparts \cite{ananthasuresh1994methodical,frecker1997topological,sigmund1997design}. Since they are monolithic designs, they require lower assembly and manufacturing cost and by comprising fewer parts and interfaces, they have comparatively less frictional, wear and tear losses. However, designing CMs is challenging, particularly in case of design-dependent loading. Therefore, dedicated TO approaches are desired.

\begin{figure*}[!ht]
	\begin{subfigure}[t]{0.45\textwidth}
		\centering
			\includegraphics[scale = 1]{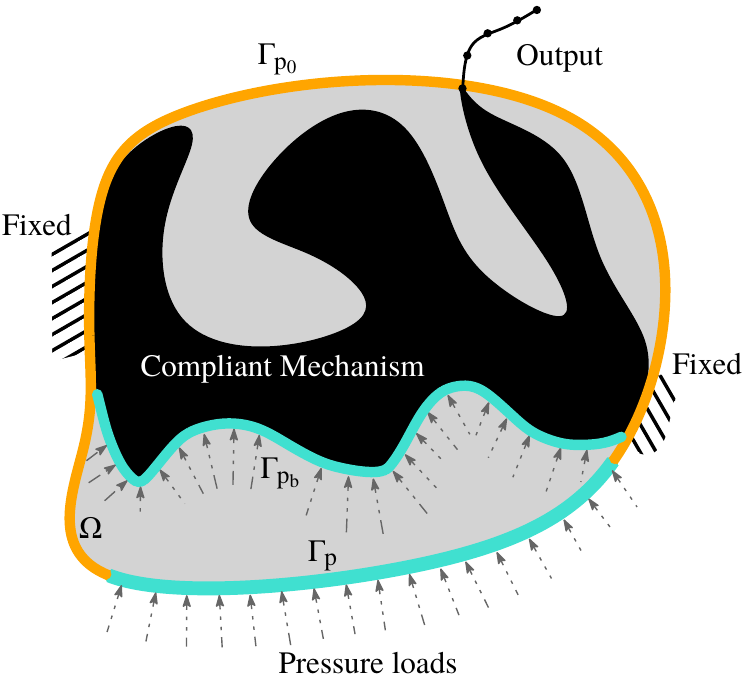}
		\caption{Compliant mechanism}
		\label{fig:voidregion}
	\end{subfigure}
	\begin{subfigure}[t]{0.45\textwidth}
		\centering
		\includegraphics[scale = 1]{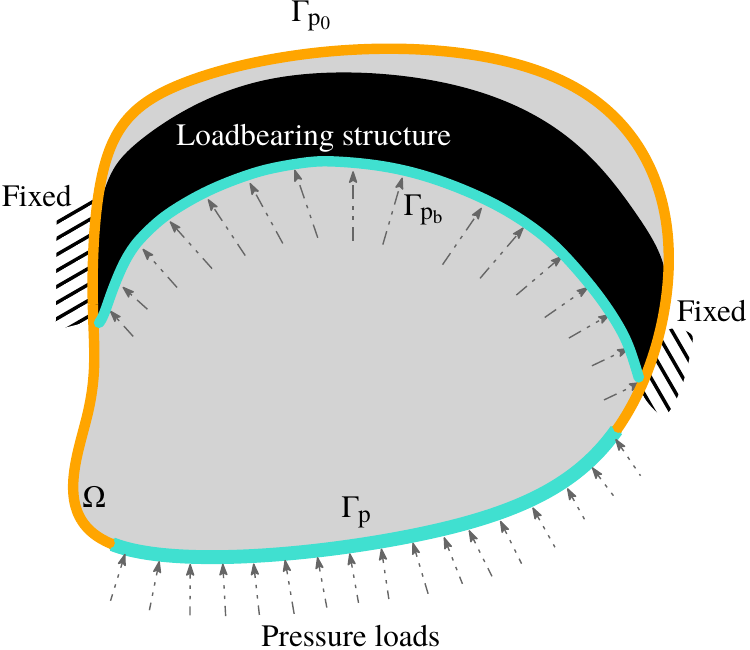}
		\caption{Loadbearing structure}
		\label{fig:voidthinsolid}
	\end{subfigure}
	\caption{Schematic representation of design domains $\mathrm{\Omega}$ for finding a pressure-actuated (depicted by gray dash-dotted arrows) optimized compliant mechanism (black solid continuum) and pressure-loaded structure (black solid continuum) in 2D. $\mathrm{\Gamma_\mathrm{p_0}}$ and $\mathrm{\Gamma_\mathrm{p}}$ boundaries (surfaces) indicate surfaces with zero and pressure loads. $\mathrm{\Gamma_\mathrm{p_b}}$ is the curve where the pressure loads are applied in the optimized designs}	\label{fig:Schematics}
\end{figure*}

To design a CM using TO, in general, an objective stemming from a flexibility measure (e.g., output/desired deformation) and a load bearing characteristic (e.g., strain-energy, stiffness, input displacement constraints and/or stress constraints) is optimized \cite{saxena2000optimal}.  The associated design domain is described using finite elements (FEs), and in a typical density-based TO method, each FE is associated with a design variable $\rho \in [0,\,1]$, which is considered herein \cite{sigmund2013topology}. $\rho=1$ indicates solid phase of an FE, whereas its void state is represented via $\rho=0$. Various applications of such mechanisms designed via TO in the case of design-independent loads can be found in  \cite{saxena2001topology,pedersen2001topology,kumar2019compliant,kumar2019computational,kumar2020topologybio,kumar2020topologyshape} and references therein. However, in case of design-dependent loading different approaches are required.  Figure~\ref{fig:Schematics} illustrates schematic design problems for a pressure-actuated CM and a pressure-loaded structure. For clarity of presentation, these are shown in 2D. A key characteristic of the problems is that the loaded surface is not predefined, but subject to change during the TO design process. Accurate calculation of load sensitivities is therefore important for these problems.

Hammer and Olhoff \cite{Hammer2000} were first to conceptualize design-dependent  pressure-loaded\footnote{For the sake of simplicity, we write only \textit{pressure load(s)} instead of \textit{design-dependent pressure load(s)}, henceforward, in this paper .} 2D structures in a TO framework. They proposed a method based on iso-density curves/surfaces which are determined using a user-defined density threshold. Du and Olhoff \cite{du2004topologicala} modified the iso-density formulation and also were first to extend the method towards designing 3D pressure-loaded structures~\cite{du2004topologicalb}. Fuchs and Shemesh \cite{Fuchs2004} used a set of variables to define  the pressure-loaded boundary explicitly, in addition to the design variables, and they also optimized pressure load variables during optimization. For an overview of 2D pressure-loaded TO approaches for designing structures and/or CMs, we refer to our recent paper on this subject \cite{kumar2020topology}.

Locating well-defined surfaces for applying pressure loads, relating pressure loads to the design vector and evaluating consistent nodal forces and their sensitivities with respect to the design vector are the central issues when considering design-dependent pressure loads in a TO setting. Compared with 2D design problems, providing a suitable solution to these challenges becomes even more complicated and involved for 3D design problems. In addition, difficulties associated with designing CMs using TO \cite{ananthasuresh1994methodical} contribute further to the above-mentioned challenges. Only few approaches  are available in the literature for 3D TO problems involving pressure loads \cite{du2004topologicalb,zhang2010topology,yang2005evolutionary,Sigmund2007,Panganiban2010}. Du and Olhoff \cite{du2004topologicalb} divided a 3D domain into a set of parallel 2D sections using a group of parallel planes to locate valid loading curves using their earlier 2D method\cite{du2004topologicala}. Thereafter, they combined all these valid loading curves to determine the appropriate surface to construct the pressure loads for the 3D problem.  A finite difference method was employed for the load sensitivities calculation, which is computationally expensive. The boundary identification scheme presented by Zhang et al. \cite{zhang2010topology} is based on an \textit{a priori} selected density threshold value. Similar to the approach by Du and Olhoff\cite{du2004topologicalb}, a 3D problem is first transformed into series of 2D problems using a group of parallel planes to determine valid loading surface. The loading surface is constructed by using the facets of FEs, and the load sensitivities are not accounted for in the approach. Steps employed in Refs. \cite{du2004topologicalb,zhang2010topology} for determining pressure loading surfaces may not be efficient and economical specially for the large-scale 3D design problems. Yang et al. \cite{yang2005evolutionary} used the ESO/BESO method in their approach. Sigmund and Clausen \cite{Sigmund2007} employed a displacement-pressure based mixed-finite element method and the three-phase material  definition (solid, void, fluid) in their approach. They demonstrated their method by optimizing pressure-loaded 2D and 3D structures. FEs used in the mixed-finite element methods have to fulfill an additional Babuska-Brezzi condition for stability \cite{zienkiewicz2005finite}. Panganiban et al. \cite{Panganiban2010} proposed an approach using a non-trivial FE formulation in association with a three-phase  material definition. They demonstrated their approach by designing a pressure-actuated 3D CM in addition to designing pressure-loaded 3D structures.

 In order to combine effectiveness in 3D CM designs under pressure loads, ease of implementation and accuracy of load sensitivities, we herein extend the method presented by Kumar et al. \cite{kumar2020topology} to 3D design problems involving both structures and mechanisms. With this, we confirm the expectation expressed in our earlier study, that the method can be naturally extended to 3D. The approach employs Darcy's law with a drainage term to identify loading surfaces (boundaries) and relates the applied pressure loads with the design vector $\bm{\rho}$. The design approach solves one additional PDE for pressure field calculation using the standard FE method. Because this involves a scalar pressure field, the computational cost is considerably lower than that of the structural analysis. The pressure field is further transformed to consistent nodal loads by considering the force originating due to pressure differences as a body force. Thus pressure forces are projected over onto a volume rather than a boundary surface, but due to the Saint-Venant's principle this difference is not relevant when evaluating global structural performance. Note that this force projection is conceptually aligned with the diffuse boundary representation commonly applied in density-based TO methods. The load sensitivities are evaluated using the adjoint-variable method. For designing loadbearing structures, compliance is minimized, whereas a multi-criteria objective \cite{saxena2000optimal} is minimized for CMs.

The layout of the paper introduce 3D method as follows. Sec.~\ref{Sec:ModelingFluidicPressureLoads} presents the proposed  pressure loading formulation in a 3D setting and  the transformation of pressure field to consistent nodal loads. A 3D test problem is also discussed for indicating the role of the drainage term in the presented approach as well as the influence of other problem parameters. The considered topology optimization problem definitions with the associated sensitivity analysis are introduced in Sec.~\ref{Sec:ProblemFormulation}.  Sec.~\ref{Sec:NumericalExamplesandDiscussion}  subsequently presents several design problems in 3D settings, including loadbearing structures and compliant mechanisms and their optimized continua. Lastly, conclusions are  drawn in Sec.~\ref{Sec:Closure}. 
	\section{Modeling of Design-dependent pressure loads}\label{Sec:ModelingFluidicPressureLoads}
In a TO setting, to determine the optimized design of a given problem, the material layout of the associated design domain $\Omega$ evolves with the optimization iterations. Consequently, in the beginning of the optimization with design-dependent loads, it may be difficult to locate a valid loading surface where such forces can be applied. In this section, we present a 3D FE modeling approach to determine a pressure field as a function of the design vector $\bm{\rho}$ using the Darcy law, which allows locating the loading surfaces implicitly. Evaluation of the consistent  nodal loads from the obtained pressure field is also described. 
 
\subsection{Concept}

In this subsection, first the Darcy-based pressure projection formulation is summarized, following our earlier 2D paper\cite{kumar2020topology}. The Darcy law which determines a pressure field through a porous medium is employed. The fluidic Darcy flux $\bm{q}$ in terms of the pressure gradient $\nabla p$, the permeability $\kappa$ of the medium, and the fluid viscosity $\mu$ can be written as
\begin{equation}\label{Eq:DarcyLaw}
\bm{q} = -\frac{\kappa}{\mu}\;\nabla p \quad = -K \;\nabla p,
\end{equation}
where $K$ is called the flow coefficient which defines the ability of a porous medium to permit fluid flow. In a density-based TO setting, each FE is characterized by a density variable that interpolates its material properties between those of the solid or void phase. Then it is natural to represent the flow coefficient of an FE with index~$i$ in terms of its filtered (physical) material density $\tilde{\rho}_i$~\cite{bourdin2001filters} and the flow coefficients of its void and solid phases such that it has a smooth variation within the design domain. Herein, we define $K(\tilde{\rho}_i)$ as
\begin{equation}\label{Eq:Flowcoefficient}
K(\tilde{\rho_i})    = K_\text{v} \left(1- (1-\epsilon) H_k(\tilde{\rho_i},\,\eta_k,\,\beta_k)\right),
\end{equation}
where
\begin{equation}\label{Eq:Heaviside}
	H_k(\tilde{\rho_i},\eta_k,\,\beta_k) = \left(\frac{\tanh{\left(\beta_k\eta_k\right)}+\tanh{\left(\beta_k(\tilde{\rho_i} - \eta_k)\right)}}{\tanh{\left(\beta_k \eta_k\right)}+\tanh{\left(\beta_k(1 - \eta_k)\right)}}\right),
\end{equation}
is the smooth Heaviside function. $\eta_k$ and $\beta_k$ are parameters which control the position of the step and the slope of $K(\tilde{\rho_i})$, respectively. Further, $\frac{K_\text{s}}{K_\text{v}} = \epsilon$ is termed  flow contrast which is set to $10^{-7}$ as motivated in Appendix \ref{Sec:Flowcontrast}, where $K_\text{v}$ and $K_\text{s}$ represent flow coefficients for void and solid elements, respectively.

As topology optimization progresses, it is expected that the pressure gradient should get confined within the solid FEs directly exposed to the pressure loading. This cannot be achieved using Eq.~\ref{Eq:DarcyLaw} only (see Sec.~\ref{Sec:qualifyingtheapp}), as it tends to distribute the pressure drop throughout the domain. Therefore, we conceptualize a volumetric drainage quantity $Q_\text{drain}$ to smoothly drain out the pressure (fluid) from the solid FEs downstream of the exposed surface. It is defined in terms of the drainage coefficient $D(\tilde{\rho_i})$, the pressure field $p$, and the external pressure $p_\text{ext}$ as
\begin{equation}\label{Eq:Drainageterm}
{Q}_\text{drain} = -D(\tilde{\rho_i}) (p - p_{\text{ext}}).
\end{equation}
The drainage coefficient $D(\tilde{\rho_i})$ is determined using a smooth Heaviside function such that pressure drops to zero for an FE with $\tilde{\rho_e} = 1$ as
\begin{equation}\label{Eq:Drainagecoefficient}
D(\tilde{\rho_i})    = \text{d}_{\text{s}}\,H_\text{d}(\tilde{\rho_i},\,\eta_\text{d},\,\beta_\text{d}),
\end{equation}
where $\beta_\text{d}$ and $\eta_\text{d}$ are the adjustable parameters, and $H_\text{d}(\tilde{\rho_i},\eta_\text{d},\,\beta_\text{d})$ is defined analogous to  Eq.~\ref{Eq:Heaviside}. The drainage coefficient of a solid FE, $\text{d}_\text{s}$, is used to control the thickness of the pressure-penetration layer and is related to $K_\text{s}$  as \cite{kumar2020topology}
\begin{equation}\label{Eq:hsrelation}
\text{d}_\text{s} =\left(\frac{\ln{r}}{\Delta s}\right)^2 K_\text{s},
\end{equation} 
where $r$ is the ratio of input pressure at depth $\Delta$s, i.e., $p|_{\Delta s} = rp_\text{in}$. Further, $\Delta s$, the penetration depth for the pressure field, can be set equal to the width or height of few FEs. This additional drainage term ensures controlled localization of the  pressure drop at the exposed structural boundary.

\subsection{3D formulation}

\begin{figure}[h!]
	\centering
	\includegraphics[scale = 1.0]{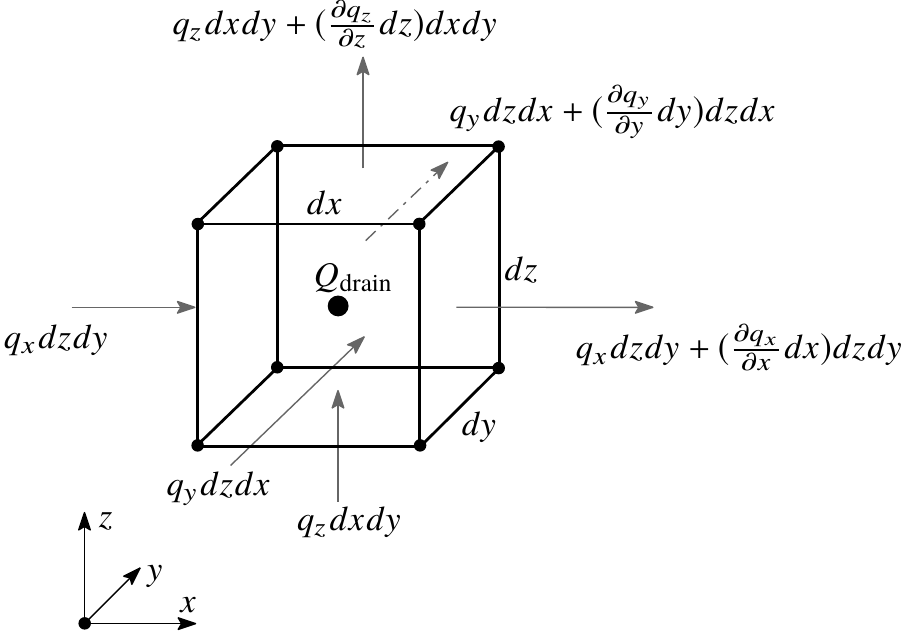}
	\caption{A schematic diagram for in- and outflow through an infinitesimal element with volume $dV = dx dy dz$. $Q_\mathrm{drain}$ is the volumetric drainage term.}
	\label{fig:Schematicsinoutflow}
\end{figure}

This section presents the 3D FE formulation for the pressure field and corresponding consistent nodal loads. The basic state equilibrium equation for the incompressible fluid flow with a drainage term can be written as (Fig.~\ref{fig:Schematicsinoutflow})

\begin{equation} \label{Eq:Basicequation}
\begin{rcases}
\begin{aligned}
\left(q_{x}dyd z\;+\; q_{y}dzdx \;+\, q_{z}d xdy\;+\,{Q}_\mathrm{drain}dV \right) &=
\left(q_{x}dydz \;+\; q_{y}dz dx\;+\, q_{z}d xdy \;+\; \left(\frac{\partial q_x}{\partial x} + \frac{\partial q_y}{\partial y} + \frac{\partial q_z}{\partial z}\right) d V\right)  \\ \text{or,}\,\,
\frac{\partial q_x}{\partial x} + \frac{\partial q_y}{\partial y}+ \frac{\partial q_z}{\partial z}-{Q}_\mathrm{drain}  = &0,\\ \text{or,}\,\,
\nabla\cdot\bm{q} -{Q}_\mathrm{drain} = &0.
\end{aligned}
\end{rcases}
\end{equation}

In view of Eqs. \eqref{Eq:DarcyLaw} and \eqref{Eq:Drainageterm}, the discretized weak form of Eq.~\eqref{Eq:Basicequation} in an elemental form   gives
\begin{equation} \label{Eq:PDEsolutionpressure}
\begin{aligned}
\underbrace{\int_{\mathrm{\Omega}_e}\left( K~ \trr{\matr{B}}_p \matr{B}_p   + D ~\trr{\matr{N}_p} \matr{N}_p \right)d {\mathrm{\Omega}_e}}_{\matr{A}_e}~\matr{p}_e =
\underbrace{\int_{\mathrm{\Omega}_e}~D~\trr{\matr{N}}_p p_\text{ext} ~~d {\mathrm{\Omega}_e} -
	\int_{\mathrm{\Gamma}_e}~ \trr{\matr{N}}_p \bm{q}_\mathrm{\Gamma} \cdot \bm{n}_e~~d {\mathrm{\Gamma}_e}}_{\matr{f}_e}
\end{aligned}
\end{equation}
where, $\matr{B}_p =\nabla\matr{N}_p$, $\bm{q}_\mathrm{\Gamma}$ represents the Darcy flux through the surface $\mathrm{\Gamma}_e$ and  $\mathbf{N}_{p} = [N_1,\, N_2,\,N_3,\,\cdots,\,N_8]$ are the shape functions for the trilinear hexahedral elements \cite{zienkiewicz2005finite} used in this paper. For other FEs, Eq.~\eqref{Eq:PDEsolutionpressure} holds similarly with different $\matr{N}_p$. In a global sense, Eq. \eqref{Eq:PDEsolutionpressure} yields 
\begin{equation}\label{Eq:Globalpressureequation}
\matr{A}\matr{p} = \matr{f},
\end{equation}
where $\matr{A}$, $\matr{p}$ and $\matr{f}$ are the global flow matrix, pressure vector and loading vector, respectively, obtained by assembling their respective elemental terms $\matr{A}_e$, $\matr{p}_e$ and $\matr{f}_e$. As  $p_\text{ext}= 0$ and $q_\mathrm{\Gamma}=0$ are assumed in this work, it follows that $\matr{f} = \mb{0}$ which leads to  $\matr{A}\matr{p} = \mb{0}$, which is solved with an appropriate input pressure $p_\text{in}$ boundary condition at a given pressure inlet surface.

The obtained pressure field is transformed to a consistent nodal force as \cite{kumar2020topology}
\begin{equation}\label{Eq:Forcepressureconversion}
\matr{F}^e = - \int_{\mathrm{\Omega}_e} \trr{\matr{N}}_\matr{u}\nabla p d {\mathrm{\Omega}_e} = - \underbrace{\int_{\mathrm{\Omega}_e} \trr{\matr{N}}_\matr{u} \matr{B}_p  d {\mathrm{\Omega}_e}}_{\matr{D}_e} \matr{p}_e,
\end{equation}
where $\mathbf{N}_\mathbf{u} = [N_1\mathbf{I},\, N_2\mathbf{I},\,N_3\mathbf{I},\,\cdots,\,N_8\mathbf{I}]$ with $\mathbf{I}$ as the identity matrix in $\mathcal{R}^3$, and  $\matr{D}_e$ representing the elemental transformation matrix. One evaluates the global nodal loads $\matr{F}$ using the following equation
\begin{equation}\label{Eq:GloablForcepressureconversion}
\matr{F} = -\matr{D}\matr{p},
\end{equation}
where  $\matr{D}$ is the global transformation matrix which is independent of the design vector. In summary, the pressure load calculation involves the following 3 main steps:
\begin{enumerate}
	\item Assembly of $\matr{A}$, which involves $K(\tilde{\bm{\rho}})$ and $D(\tilde{\bm{\rho}})$ as design-dependent terms (Eqs.~\ref{Eq:Flowcoefficient},\,\ref{Eq:Heaviside},\,\ref{Eq:Drainagecoefficient},\,\ref{Eq:PDEsolutionpressure}) 
	\item Solve $\matr{A}\matr{p} = \matr{0}$ (Eq.~\ref{Eq:Globalpressureequation})
	\item Calculation of $\matr{F} = -\matr{D}\matr{p}$ (Eq.~\ref{Eq:GloablForcepressureconversion})
\end{enumerate}
Note that step 2 involves a linear system with three times fewer degrees of freedom compared to the structural problem, as each node only has a single pressure state. Hence in terms of computational cost, the structural analysis remains dominant.

\subsection{Qualifying the approach}\label{Sec:qualifyingtheapp}
This section presents a test problem for illustrating the method and demonstrating the importance of the drainage term (Eq.~\ref{Eq:Drainageterm}) in the presented approach. An additional test problem  is included in Appendix~\ref{Sec:Flowcontrast} to study the effect of flow contrast $\epsilon$.

\begin{figure}[!h]
	\begin{subfigure}[t]{0.32\textwidth}
		\centering
		\includegraphics[width=\linewidth]{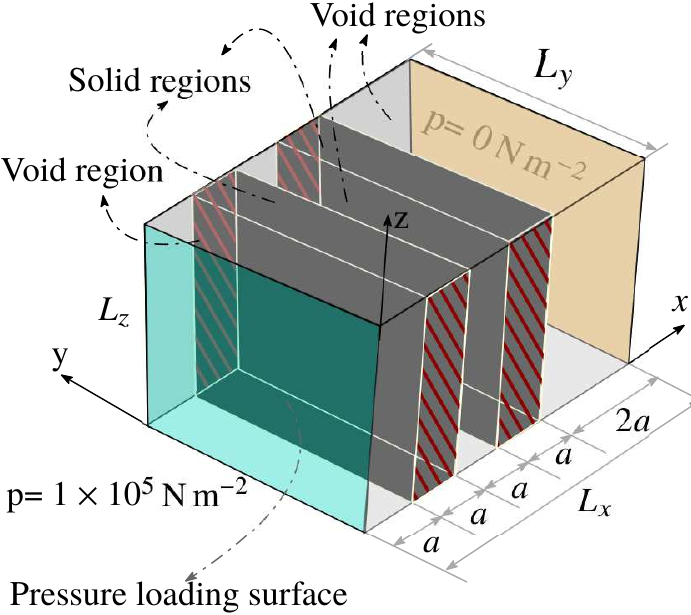}
		\caption{}
		\label{fig:test3D}
	\end{subfigure}
	\begin{subfigure}[t]{0.32\textwidth}
		\centering
		\includegraphics[width=\linewidth]{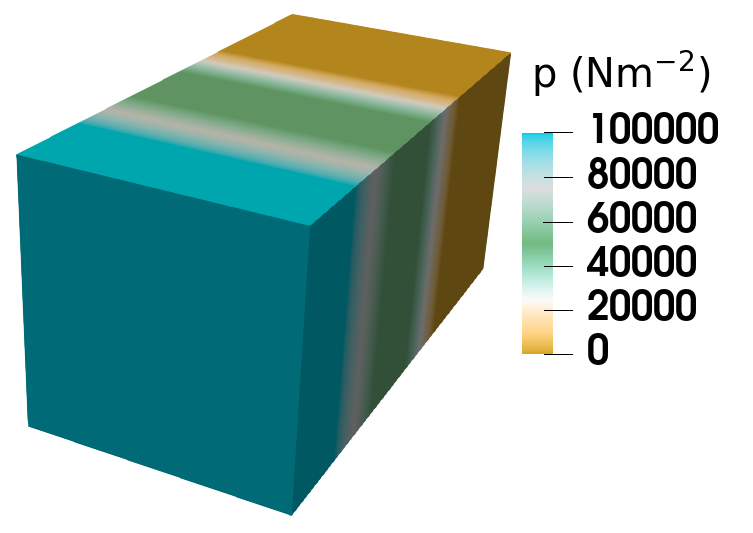} 
		\caption{}
		\label{fig:PvariationWD}
	\end{subfigure}
	\begin{subfigure}[t]{0.32\textwidth}
		\centering
		\includegraphics[width=0.925\linewidth]{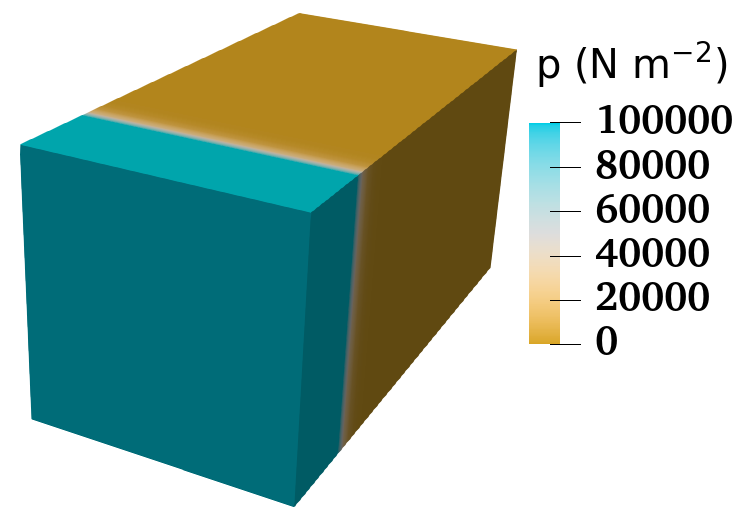}  %
		\caption{}
		\label{fig:PvariationWithD}
	\end{subfigure}
	\caption{(\subref{fig:test3D}) Design domain specification to show importance of the drainage term. (\subref{fig:PvariationWD}) Pressure field variation ($\si{\newton\per\meter\squared}$) without drainage term, (\subref{fig:PvariationWithD}) Pressure field variation ($\si{\newton\per\meter\squared}$) with drainage term. Fixed planes are hatched in (\subref{fig:test3D}).}\label{fig:Test1Study}
\end{figure}

Figure~\ref{fig:test3D} depicts the design specifications of the test problem. We consider a domain of $L_x\times L_y \times L_z$ = $ 0.02\times 0.01\times 0.01$ \si{\meter\cubed}, with a pressure load of \SI{1e5}{\newton\per\meter\squared} is applied on the front face of the domain, and zero pressure on the rear face. The total normal force experienced by the front face is $F_x =100000\times0.01\times 0.01\, \text{N}= 10$ N. The domain has two solid regions  with dimensions  $\frac{L_x}{6}\times L_y\times L_z\si{\meter\cubed}$, which are separated by $\frac{L_x}{6}$ (Fig.~\ref{fig:Test1Study}). The remaining regions of the domain are considered void. The design domain is discretized using $48\times 24\times 24$ trilinear hexahedral FEs. The other required parameters are same as those mentioned in  Table~\ref{Table:T1}.

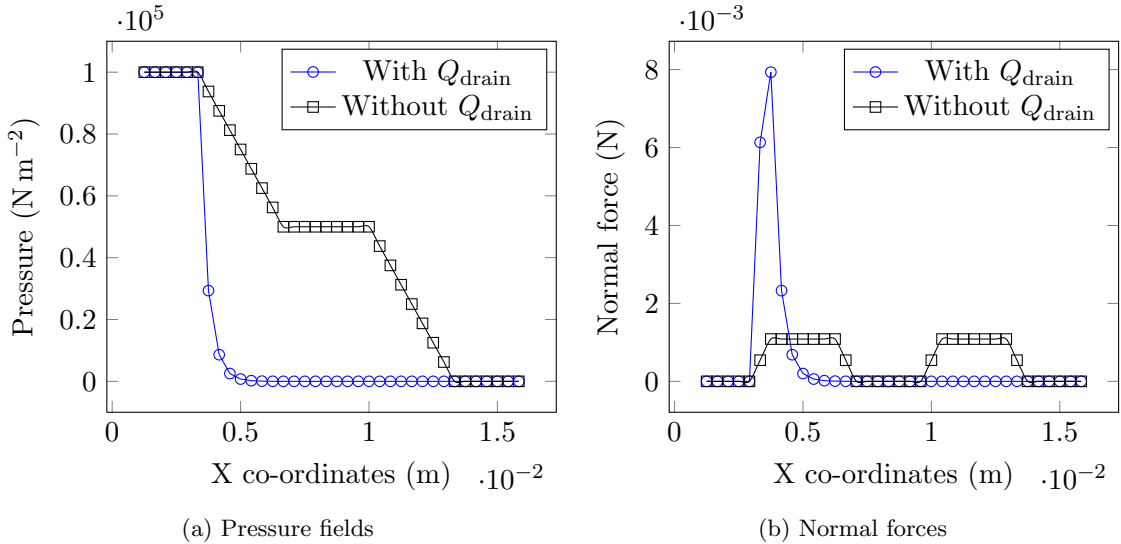
\begin{figure}[h!]
	\begin{subfigure}[t]{0.5\textwidth}
		\centering
		\begin{tikzpicture} 	
		\pgfplotsset{compat = 1.3}
		\begin{axis}[
		width = 1\textwidth,
		xlabel=X co-ordinates (m),
		ylabel= Pressure ($\si{\newton\per\meter\squared}$)]
		\pgfplotstableread{3DpressureWithD.txt}\mydata;
		\addplot[mark=o,blue]
		table {\mydata};
		\addlegendentry{With $Q_\text{drain}$}
		\pgfplotstableread{3DpressureWD.txt}\mydata;
		\addplot[smooth,mark=square,black]
		table {\mydata};
		\addlegendentry{Without $Q_\text{drain}$}
		\end{axis}
		\end{tikzpicture}
		\caption{Pressure fields}
		\label{fig:3Dpressurevariation}
	\end{subfigure}
	\begin{subfigure}[t]{0.5\textwidth}
		\centering
		\begin{tikzpicture} 	
		\pgfplotsset{compat = 1.3}
		\begin{axis}[
		width = 1\textwidth,
		xlabel=X co-ordinates (m),
		ylabel= Normal force (N)]
		\pgfplotstableread{3DforceWithD.txt}\mydata;
		\addplot[mark=o,blue]
		table {\mydata};
		\addlegendentry{With $Q_\text{drain}$}
		\pgfplotstableread{3DforceWD.txt}\mydata;
		\addplot[smooth,mark=square,black]
		table {\mydata};
		\addlegendentry{Without $Q_\text{drain}$}
		\end{axis}
		\end{tikzpicture}
		\caption{Normal forces}
		\label{fig:3Dforcevariation}
	\end{subfigure}
	\caption{The pressure field and respective nodal force variations along the center $x-$axis with and without drainage term are depicted. One notices a smooth variation with $Q_\text{drain}$, whereas without the drainage term as expected the pressure field shows a step variation over two solid regions.}\label{fig:3Dpressforcevariation}
\end{figure}

Figure~\ref{fig:PvariationWD} and  Fig.~\ref{fig:PvariationWithD} depict the pressure field variation within the domain with and without drainage term (Eq.~\ref{Eq:Globalpressureequation}). The pressure and nodal force variations along the center of the domain in the $x-$axis are depicted with and without $Q_\text{drain}$ in Fig.~\ref{fig:3Dpressforcevariation}. One notices that if the drainage term is not considered, the pressure gradient does not get confined as soon as the pressure loading faces the first solid region (Figs.~\ref{fig:PvariationWD} and \ref{fig:3Dpressurevariation}), which is undesirable for the intended purpose. However, a correct behavior is seen when including the drainage term (Figs.~\ref{fig:PvariationWithD} and \ref{fig:3Dforcevariation}). The corresponding nodal force variations are also reported in Fig.~\ref{fig:3Dpressforcevariation}. It is found that the total normal force experienced by the design in Fig.~\ref{fig:PvariationWithD} is  $\SI{10}{\newton}$ which is equal to the original force applied (Fig.~\ref{fig:test3D}).  

	\section{Problem formulation}\label{Sec:ProblemFormulation}
This section presents the optimization problem formulation and the sensitivities of the objectives with respect to the design vector $\bm{\rho}$ using the adjoint-variable method. 
\subsection{Optimization formulation}\label{SubSec:OptimizationFormulation}

The optimization problem is formulated using a density-based TO framework, wherein each FE is associated with a design variable~$\rho$ which is further filtered using the classical density filter \cite{bourdin2001filters}. The filtered design variable $\tilde{\rho_i}$ is evaluated as the weighted average of the design variable $\rho_j$  as \cite{bourdin2001filters}
\begin{equation}\label{Eq:densityfilter}
\tilde{\rho_i} = \frac{\sum_{j=1}^{N_e} v_j \rho_j w(\mathbf{x}_j)}{\sum_{j=1}^{N_e} v_j w(\mathbf{x}_j)} ,
\end{equation}
where $N_e$ represents the total number of FEs lie within the filter radius $R_\text{fil}$ for the $i^\text{th}$ element, $v_j$ is the volume of the $j^\text{th}$ element  and $w(\matr{x}_j)$, the weight function, is defined as
\begin{equation}\label{Eq:weightdensity}
w(\matr{x}_j)=\max\left(0,\,1-\frac{||\matr{x}_i -\matr{x}_j||}{R_\text{fill}}\right),
\end{equation} 
where $||\matr{x}_i -\matr{x}_j||$ is the Euclidean distance between the $i^\text{th}$ and $j^\text{th}$ FEs.  $\matr{x}_i$ and $\matr{x}_j$ indicate the center coordinates of the $i^\text{th}$ and $j^\text{th}$ FEs, respectively. The derivative of filtered density with respect to  the design variable can be evaluated as
\begin{equation}\label{Eq:derivativefilteractual}
\pd{\tilde{\rho_i}}{\rho_j} = \frac{ v_j w(\matr{x}_j)}{\sum_{k=1}^{N_e} v_k w(\matr{x}_k)}.
\end{equation}
The Young's modulus of each FE is evaluated using the modified SIMP (Solid Isotropic Material with Penalization) formulation as
\begin{equation}\label{Eq:SIMP}
E_e(\rho_e) = E_0 + \tilde{\rho_e}^\zeta (E_1 - E_0), \qquad \tilde{\rho_e}\in[0, \, 1]
\end{equation}
where, $E_1$ is the Young's modulus of the actual material, $E_0 = 10^{-6}E_1$ is set, and the penalization parameter $\zeta$ is set to 3, which guides the TO towards \textquotedblleft 0-1\textquotedblright\, solutions. 

The following topology optimization problem is solved:
 \begin{equation}\label{Eq:Optimizationequation}
\begin{rcases}
& \underset{\bm{\rho}}{\text{min}}
& &{f_0}\\
& \text{such that:}  &&\,\, \mathbf{Ap} = \mathbf{0 }\\
&  &&\,\,\mathbf{Ku = F} = -\mathbf{D p}\\
&  &&\,\,\mathbf{Kv = F_\mathrm{d}}\\
&  && \,\,\text{g}_1=\frac{ V(\bm{\tilde{\rho}})}{V^*}-1\le 0\\
&  && \mathbf{0}\le\bm{\rho}\le \mathbf{1}
\end{rcases},
\end{equation}
where $f_0$ is the objective function to be optimized. The global stiffness matrix and displacement vector are denoted by $\mathbf{K}$ and $\mathbf{u}$, respectively. For designing  pressure-loaded 3D loadbearing structures, compliance, i.e., $ f_0 = 2SE$ is minimized\footnote{For loadbearing structure designs, $\mathbf{Kv = F_\mathrm{d}}$ is not considered}, whereas for the pressure-actuated 3D compliant mechanism designs a multi-criteria \cite{saxena2000optimal} objective, i.e., $f_0 = -\mu\frac{MSE}{SE}$ is minimized. $SE$ and $MSE$  represent the strain energy and mutual strain energy of the design, respectively. Further, $\mu$, a scaling factor, is employed primarily to adjust the magnitude of the  objective to suit the MMA optimizer, and $MSE =\trr{\mathbf{v}}\mathbf{Ku}$ is equal to the output deformation wherein $\mb{F}_\text{d}\,(= \mathbf{Kv)}$ is the unit dummy force applied  in the direction of the desired deformation at the output location \cite{saxena2000optimal}. Furthermore, $V$ and $V^*$ are the actual and permitted volumes of the designs, respectively.

\subsection{Sensitivity analysis} 

We use the gradient-based MMA optimizer \cite{svanberg1987method} for the topology optimization. The adjoint-variable method  is employed to determine the sensitivities\footnote{A detailed description is given in Ref. \cite{kumar2020topology}} of the objectives and constraints with respect to the design variables. One can write an aggregate performance function $\mathcal{L}$ for evaluating the sensitivities as

\begin{equation}\label{Eq:performancefunction}
\begin{aligned}
{\mathcal{L} (\mathbf{u},\mathbf{v},\,\tilde{\bm{\rho}})} = f_0(\mathbf{u},\mathbf{v},\tilde{\bm{\rho}}) + \trr{\bm{\lambda}}_1 \left(\mathbf{Ku +{D p}}\right) + \trr{\bm{\lambda}}_2 ( \mathbf{Ap}) + \trr{\bm{\lambda}}_3 (\mathbf{Kv-F_\mathrm{d}}),
\end{aligned}
\end{equation}
where  $\bm{\lambda}_1,\,\bm{\lambda}_2$ and $\bm{\lambda_3}$, the Lagrange multipliers, are determined as \cite{kumar2020topology}

\begin{equation}\label{Eq:lagrangemultipliers}
\begin{rcases}
\trr{\bm{\lambda}}_1  &= -\pd{f_0(\mathbf{u},\, \mathbf{v},\,\tilde{\bm{\rho}})}{\mathbf{u}} \inv{\mathbf{K}}\\
\trr{\bm{\lambda}}_2  & = -\trr{\bm{\lambda}}_1 \mathbf{D}\inv{\mathbf{A}}\\
\trr{\bm{\lambda}}_3  &= -\pd{f_0(\mathbf{u},\, \mathbf{v},\,\tilde{\bm{\rho}})}{\mathbf{v}} \inv{\mathbf{K}}
\end{rcases}.
\end{equation}
For the loadbearing structures and CMs, Eq.~\ref{Eq:lagrangemultipliers} yields

\begin{equation}\label{Eq:lagrangemultipliersstructure}
\trr{\bm{\lambda}}_1  = -2\trr{\mathbf{u}},\,
\trr{\bm{\lambda}}_2   =  2\trr{\mathbf{u}} \mathbf{D}\inv{\mathbf{A}},\,\quad \left(\text{Structure}\right);
\end{equation}
\begin{equation}\label{Eq:lagrangemultiplierCM}
\trr{\bm{\lambda}}_1  = \mu\left(\frac{\trr{\mathbf{v}}}{SE} - \trr{\mathbf{u}}\frac{MSE}{(SE)^2}\right),\,
\trr{\bm{\lambda}}_2   = -\mu\left(\frac{\trr{\mathbf{v}}}{SE} - \trr{\mathbf{u}}\frac{MSE}{(SE)^2}\right) \mathbf{D}\inv{\mathbf{A}},\,
\trr{\bm{\lambda}}_3  = \mu\frac{\trr{\mathbf{u}}}{SE},\,\quad \left(\text{CMs}\right).
\end{equation}
Now, one can  evaluate the objective sensitivities as

\begin{equation}\label{Eq:objectivesensitivity}
\frac{d f_0}{d\tilde{\bm{\rho}}} = \pd{f_0}{\tilde{\bm{\rho}}} + \trr{\bm{\lambda}}_1\pd{\mathbf{K}}{\tilde{\bm{\rho}}}\mathbf{u} + \trr{\bm{\lambda}}_2\pd{\mathbf{A}}{\tilde{\bm{\rho}}}\mathbf{p} +  \trr{\bm{\lambda}}_3\pd{\mathbf{K}}{\tilde{\bm{\rho}}}\mathbf{v},
\end{equation}
where vectors $\mathbf{u},\,\mathbf{p}$ and $\mathbf{v}$ also includes their prescribed values. Now, in view of Eqs.~\ref{Eq:Optimizationequation},  \ref{Eq:lagrangemultiplierCM} and \ref{Eq:objectivesensitivity}, one can subsequently determine the sensitivities for loadbearing structures and CMs with respect to the filtered design vector $\tilde{\bm{\rho}}$ as

 \begin{equation}\label{Eq:ComplianceSens}
\frac{d f_0}{d\tilde{\bm{\rho}}} =  -\trr{\mathbf{u}}\pd{\mathbf{K}}{\tilde{\bm{\rho}}}\mathbf{u} + \underbrace{2\trr{\mathbf{u}} \mathbf{D}\inv{\mathbf{A}}\pd{\mathbf{A}}{\tilde{\bm{\rho}}}\mathbf{p}}_{\text{Load sensitivities}},
\end{equation}
and 
\begin{equation}\label{Eq:CMSens}
\frac{d f_0}{d\tilde{\bm{\rho}}} =  \mu\left[\frac{MSE}{(SE)^2}\left(-\frac{1}{2}\trr{\mathbf{u}}\pd{\mathbf{K}}{\tilde{\bm{\rho}}}\mathbf{u}\right) + \frac{1}{SE}\left(\trr{\mathbf{u}}\pd{\mathbf{K}}{\tilde{\bm{\rho}}}\mathbf{v}\right)+
\underbrace{\frac{MSE}{(SE)^2}\left(\trr{\mathbf{u}} \mathbf{D}\inv{\mathbf{A}}\pd{\mathbf{A}}{\tilde{\bm{\rho}}}\mathbf{p}\right) + \frac{1}{SE}\left(-\trr{\mathbf{v}} \mathbf{D}\inv{\mathbf{A}}\pd{\mathbf{A}}{\tilde{\bm{\rho}}}\mathbf{p}\right)}_{\text{Load sensitivities}}\right],
\end{equation}
respectively. Further, one finds the sensitivities of the objectives with respect to the design vector $\bm{\rho}$ using Eqs.~\ref{Eq:derivativefilteractual}, ~\ref{Eq:ComplianceSens} and ~\ref{Eq:CMSens}. The load sensitivity terms for both the objectives can be readily evaluated using Eqs.\ref{Eq:ComplianceSens} and \ref{Eq:CMSens}. As the  pressure loads acting on the structure depend on the design, it is important to include these terms in the optimization.

	\section{Numerical Examples and Discussion} \label{Sec:NumericalExamplesandDiscussion}
In this section, various small deformation 3D compliant mechanisms actuated via pressure loads and 3D pressure-loaded structures are designed to demonstrate the effectiveness and versatility of the presented approach. 
\begin{table}[h!] 
	\centering
	\begin{tabular}{ l c c }
		\hline
		\hline
		\textbf{Nomenclature} & \textbf{Notation} & \textbf{Value}  \\ \hline
		Young's Modulus of a solid FE ($\tilde{\rho}=1$) & $E_1$ & $\SI{5e8}{\newton\per\square\meter}$ \rule{0pt}{3ex}\\ 
		Poisson's ratio & $\nu$ & $0.40$\\ 
		SIMP Penalization  &$\zeta$ &$3$ \rule{0pt}{3ex}   \\
		Young's Modulus of a void FE ($\tilde{\rho}=0$) & $E_0$ &$E_1 \times 10^{-6} \si{\newton\per\square\meter}$ \\
		External Move limit & $\Delta \bm{\rho}$ & 0.1 per iteration\\
		Input pressure load &$p_\mathrm{in}$ & $\SI{1e5}{\newton\per\square\meter}$ \rule{0pt}{3ex}\\
		$K(\tilde{\bm{\rho}})$ step location	& 	$\eta_k$ 	& 0.3 \rule{0pt}{3ex}\\
		$K(\tilde{\bm{\rho}})$ slope	at step	& 	$\beta_k$	& 10\\
		$D(\tilde{\bm{\rho}})$ step location	& 	$\eta_h$ 	& 0.2\\
		$D(\tilde{\bm{\rho}})$ slope	at step	& 	$\beta_h$	& 10\\ 
		Flow coefficient of a void FE	&   $k_\mathrm{v}$  & $\SI{1}{\meter\tothe{4}\per\newton\per\second}$\\
		Flow coefficient of a solid FE &   $k_\mathrm{s} $  & $K_\mathrm{v}\times\SI{e-7}{\meter\tothe{4}\per\newton\per\second}$\\
		Drainage from solid	&   $h_\mathrm{s} $  & $\left(\frac{\ln{r}}{\Delta s}\right)^2 K_\mathrm{s}$ \\
		Remainder of input pressure at $\Delta s$ &r& 0.1\\\hline\hline
	\end{tabular}
	\caption{Various parameters used in this paper.}\label{Table:T1}
\end{table}
Trilinear hexahedral FEs are employed to parameterize the design domains. Optimization parameters with their nomenclature, symbol and unit are mentioned in Table~\ref{Table:T1} and any alteration is reported in the respective problem definition. TO is performed using an in-house MATLAB code with the MMA optimizer. The maximum number of MMA iterations are set to 100 and 250 for optimizing the loadbearing structures and compliant mechanisms, respectively.  The linear systems from state and adjoint equations are solved using the conjugate gradient method in combination with incomplete Cholesky preconditioning.
 
\begin{figure}[h!]
	\begin{subfigure}[t]{0.45\textwidth}
		\centering
		\includegraphics[width=0.75\linewidth]{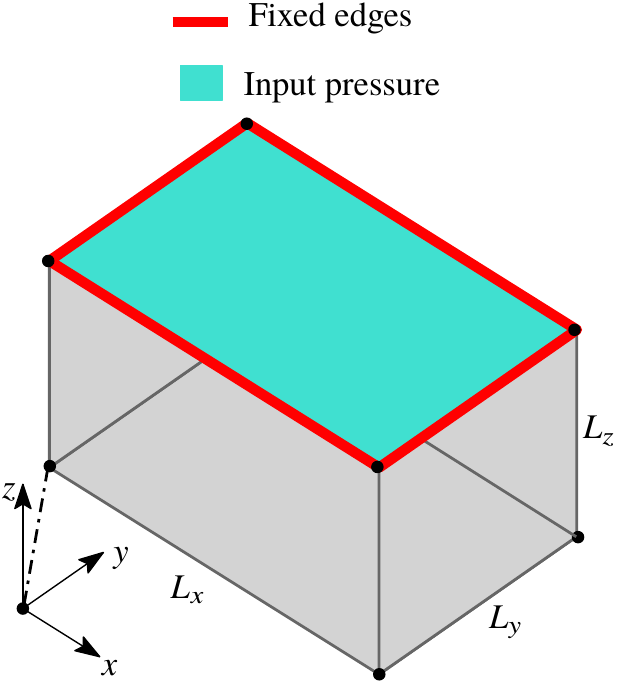}
		\caption{}
		\label{fig:DesigndomainSS_1}
	\end{subfigure}
	\begin{subfigure}[t]{0.45\textwidth}
		\centering
		\includegraphics[width=0.75\linewidth]{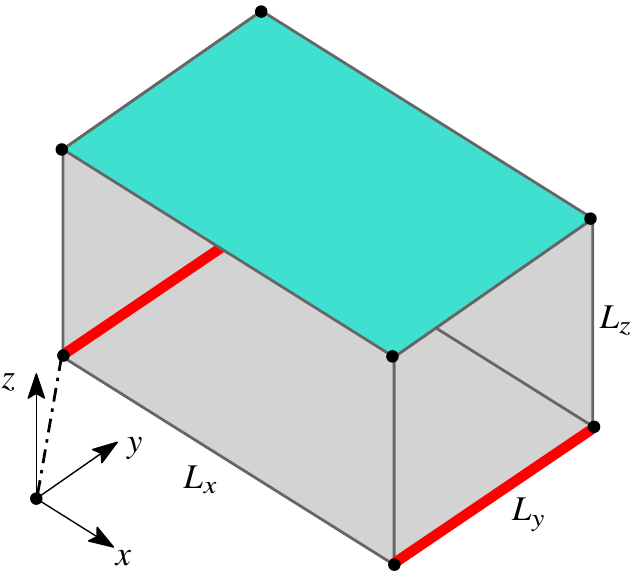}
		\caption{}
		\label{fig:DesigndomainSS_2}
	\end{subfigure}
	\caption{Design domains and problem definitions of the loadbearing structures. (\subref{fig:DesigndomainSS_1}) Lid  domain, (\subref{fig:DesigndomainSS_2}) Externally pressurized domain.}\label{fig:DesigndomainsProblemstiff}
\end{figure}

\begin{figure}[h!]
	\begin{subfigure}[t]{0.31\textwidth}
		\centering
		\includegraphics[width=\linewidth]{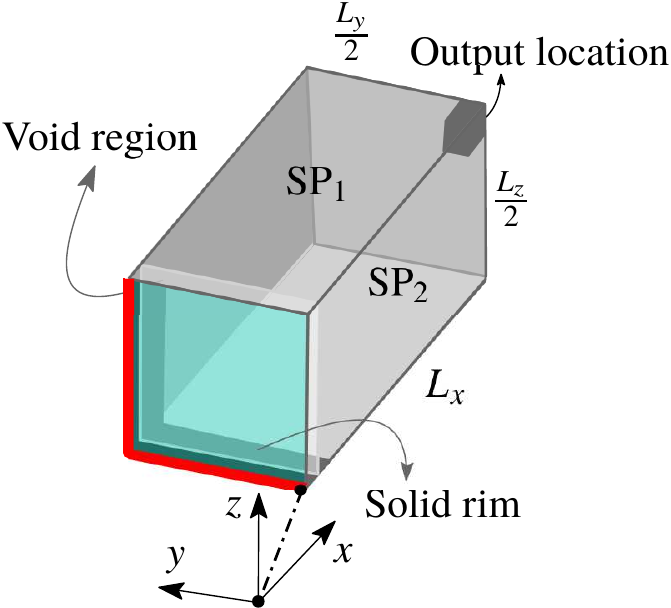}
		\caption{}
		\label{fig:DesigndomainCM_inverter}
	\end{subfigure}
	\begin{subfigure}[t]{0.31\textwidth}
		\centering
		\includegraphics[width=1\linewidth]{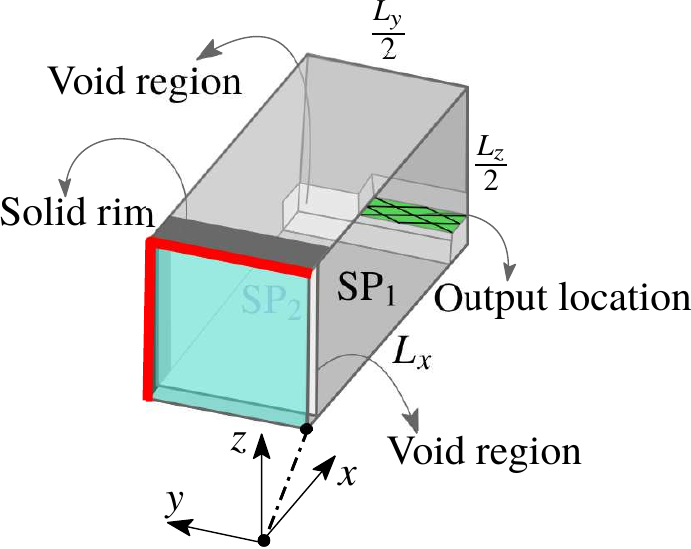}
		\caption{}
		\label{fig:DesigndomainCM_gripper}
	\end{subfigure}
	\begin{subfigure}[t]{0.31\textwidth}
		\centering
		\includegraphics[width=1\linewidth]{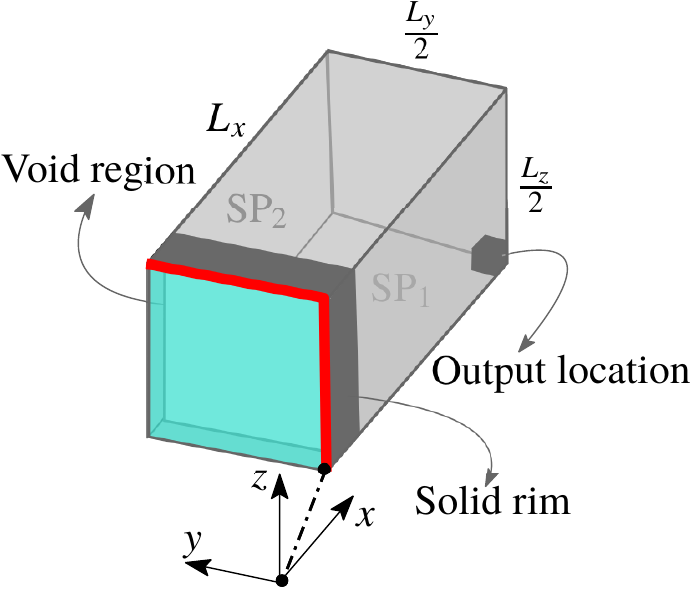}
		\caption{}
		\label{fig:DesigndomainCM_magnifier}
	\end{subfigure}
	\caption{Design domains and problem definitions of the CMs. (\subref{fig:DesigndomainCM_inverter}) Quarter inverter domain (\subref{fig:DesigndomainCM_gripper}) Quarter gripper domain, void non-design region and jaw (green) of the gripper are shown  (\subref{fig:DesigndomainCM_magnifier}) Quarter magnifier domain. Output location for each mechanism design is displayed, where springs representing stiffness of a workpiece are attached. A void non-design domain with a rim of solid non-design region around the pressure inlet area is considered for each mechanism design, representing the maximum pressure inlet geometry. SP$_1$ and SP$_2$ indicate the symmetric planes for the quarter compliant mechanism designs.}\label{fig:DesigndomainsProblemCMs}
\end{figure}
\subsection{Pressure loaded structures}
In this section, two pressure-loaded structure design optimization problems i.e., a lid (Fig.~\ref{fig:DesigndomainSS_1}) and an externally pressurized structure (Fig.~\ref{fig:DesigndomainSS_2}) are presented. The lid design problem appeared initially in the work of Du and Olhoff \cite{du2004topologicalb}, whereas the externally pressure structure design problem is taken from Zhang et al. \cite{zhang2010topology}.

Let $L_x$, $L_y$ and $L_z$ represent the design domain dimensions in $x-$, $y-$ and $z-$directions, respectively.  $L_x\times L_y \times L_z = {0.2}\times{0.1}\times\SI{0.1}{\meter\cubed}$ is considered for the lid and the externally pressurized design. An inlet pressure $p_\text{in}$ of   $\SI{1}{\bar}$ is applied on the top face of the domains. Edges depicted in red are fixed for all design domains (Fig.~\ref{fig:DesigndomainsProblemstiff}). The permitted volume fraction for each example is set to 0.25.  The lid design is optimized considering the full model\footnote{Symmetry conditions are not exploited}, so any tendency of the problem to break the symmetry can be observed. However, the externally pressurized design is optimized by exploiting one of its symmetric conditions, i.e., only half the design domain is considered. The full lid and a symmetric half externally pressurized  are parameterized by $120\times60\times60$ and $80\times80\times80$  hexahedral FEs, respectively. The filter radius is set to $r_\text{min}=\sqrt{3}\min\left(\frac{L_x}{N_{ex}},\frac{L_y}{N_{ey}},\frac{L_z}{N_{ez}}\right)$ for all the solved problems.

\begin{figure}[h!]
	\begin{subfigure}[t]{0.19\textwidth}
		\centering
		\includegraphics[width=\linewidth]{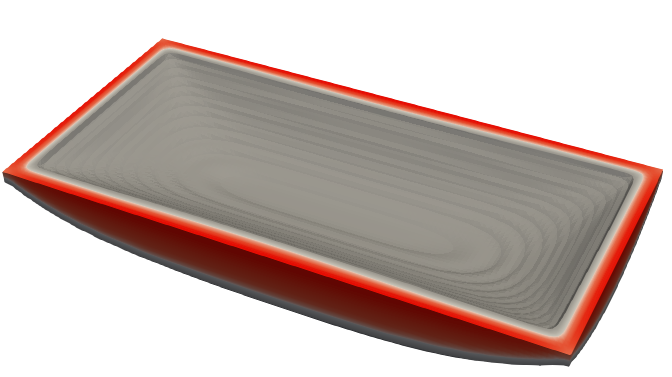}
		\caption{Arbitrary direction}
		\label{fig:Bathtubview1}
	\end{subfigure}
	\begin{subfigure}[t]{0.19\textwidth}
		\centering
		\includegraphics[width=\linewidth]{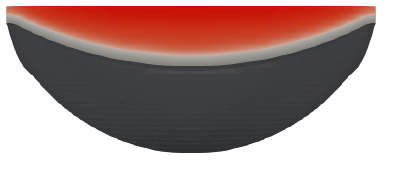}
		\caption{$+x-$direction}
		\label{fig:Bathtubview2}
	\end{subfigure}
	\begin{subfigure}[t]{0.19\textwidth}
		\centering
		\includegraphics[width=\linewidth]{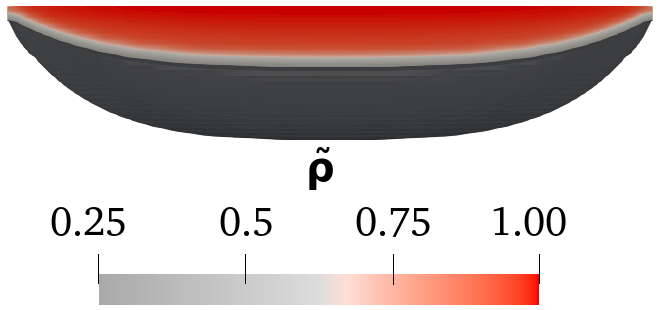}
		\caption{$+y-$direction}
		\label{fig:Bathtubview3}
	\end{subfigure}
	\begin{subfigure}[t]{0.19\textwidth}
		\centering
		\includegraphics[width=\linewidth]{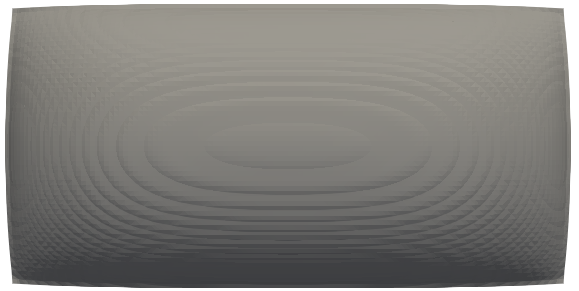}
		\caption{+$z-$direction}
		\label{fig:Bathtubview4}
	\end{subfigure}
	\begin{subfigure}[t]{0.19\textwidth}
		\centering
		\includegraphics[width=\linewidth]{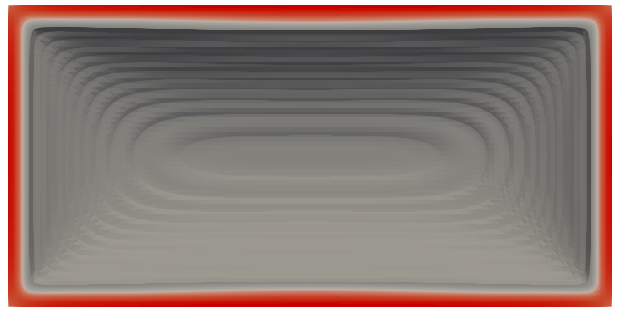}
		\caption{$-z-$direction}
		\label{fig:Bathtubview5}
	\end{subfigure}
	\caption{The optimized lid design is shown in different view directions.  The figure in \subref{fig:Ex2view2} indicates the material density color scheme which is kept same for all the solved problems.}\label{fig:Bathtubdesignresult}
\end{figure}

\begin{figure}
	\begin{subfigure}[t]{0.19\textwidth}
		\centering
		\includegraphics[width=\linewidth]{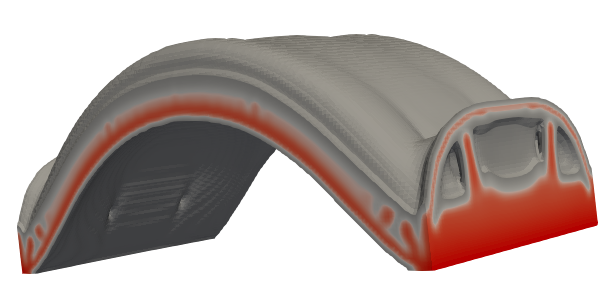}
		\caption{Arbitrary direction}
		\label{fig:Ex2view1}
	\end{subfigure}
	\begin{subfigure}[t]{0.19\textwidth}
		\centering
		\includegraphics[width=0.70\linewidth]{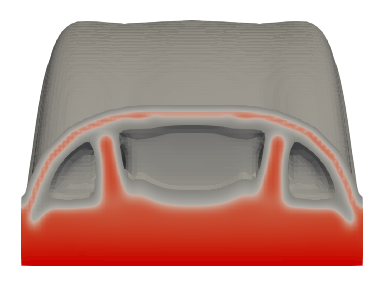}
		\caption{$+x-$direction}
		\label{fig:Ex2view2}
	\end{subfigure}
	\begin{subfigure}[t]{0.19\textwidth}
		\centering
		\includegraphics[width=\linewidth]{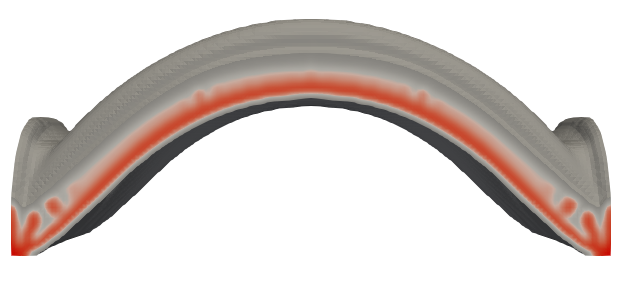}
		\caption{$+y-$direction}
		\label{fig:Ex2view3}
	\end{subfigure}
	\begin{subfigure}[t]{0.19\textwidth}
		\centering
		\includegraphics[width=\linewidth]{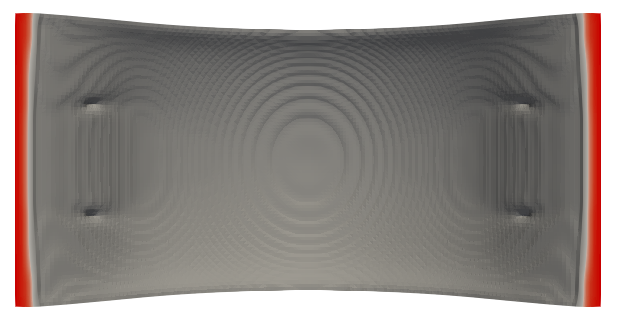}
		\caption{$+z-$direction}
		\label{fig:Ex2view4}
	\end{subfigure}
	\begin{subfigure}[t]{0.19\textwidth}
		\centering
		\includegraphics[width=\linewidth]{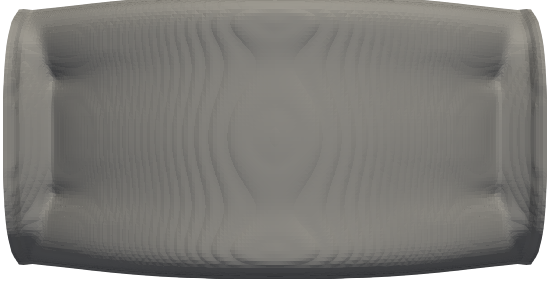}
		\caption{$-z-$direction}
		\label{fig:Ex2view5}
	\end{subfigure}
	\caption{The optimized externally pressurized design is shown in different view directions.}\label{fig:Ex2result}
\end{figure}

\begin{figure}[h!]
	\begin{subfigure}[t]{0.45\textwidth}
		\centering
		\includegraphics[width=\linewidth]{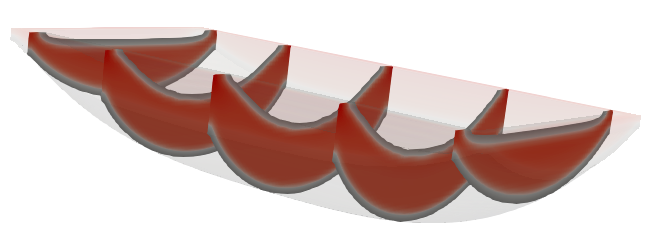}
		\caption{}
		\label{fig:EX1CSV}
	\end{subfigure}
	\begin{subfigure}[t]{0.45\textwidth}
		\centering
		\includegraphics[width=0.65\linewidth]{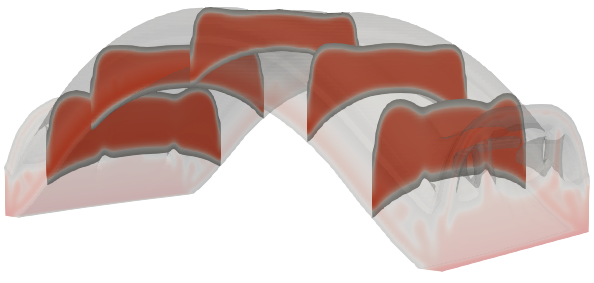}
		\caption{}
		\label{fig:EX2CSV}
	\end{subfigure}
	\caption{Material distributions of the optimized lid and externally pressurized design with respect to different cross sections in arbitrary directions are shown in \subref{fig:EX1CSV} and \subref{fig:EX2CSV}, respectively.}\label{fig:EX_stiff_CSV}
\end{figure}

\begin{figure}[h!]
	\begin{subfigure}[t]{0.5\textwidth}
		\centering
		\begin{tikzpicture} 	
		\pgfplotsset{compat = 1.3}
		\begin{axis}[
		width = 1\textwidth,
		xlabel=MMA iteration,
		ylabel= Normalized Compliance (N m)]
		\pgfplotstableread{Ex1Objective.txt}\mydata;
		\addplot[smooth,mark=square*,blue,,mark size=0.5pt]
		table {\mydata};
		\addlegendentry{Lid design}
		\pgfplotstableread{Ex2Objective.txt}\mydata;
		\addplot[smooth,mark=*,black,mark size=0.5pt]
		table {\mydata};
		\addlegendentry{Ext. pressurized design}
		\end{axis}
		\end{tikzpicture}
		\caption{}
		\label{fig:AllObjectiveplot}
	\end{subfigure}
	\quad
	\begin{subfigure}[t]{0.5\textwidth}
		\centering
		\begin{tikzpicture}
		\pgfplotsset{compat = 1.3}
		\begin{axis}[
		width = 1\textwidth,
		xlabel=MMA iteration,
		ylabel= Volume fraction,
		ytick ={0.20,0.21,0.22,0.23,0.24,0.25,0.30},legend pos=south east]
		\pgfplotstableread{Ex1Volume.txt}\mydata;
		\addplot[smooth,mark=square*,blue,,mark size=0.5pt]
		table {\mydata};
		\addlegendentry{Lid design}
		\pgfplotstableread{Ex2Volume.txt}\mydata;
		\addplot[smooth,mark=*,black,mark size=0.5pt]
		table {\mydata};
		\addlegendentry{Ext. pressurized design}
		\end{axis}
		\end{tikzpicture}
		\caption{}
		\label{fig:AllVolumeplot}
	\end{subfigure}
	\caption{Objective and volume fraction convergence plots for loadbearing structure problems. (\subref{fig:AllObjectiveplot}) Compliance history, and (\subref{fig:AllVolumeplot}) Volume fraction history.}
	\label{fig:Example1Convergence}
\end{figure}
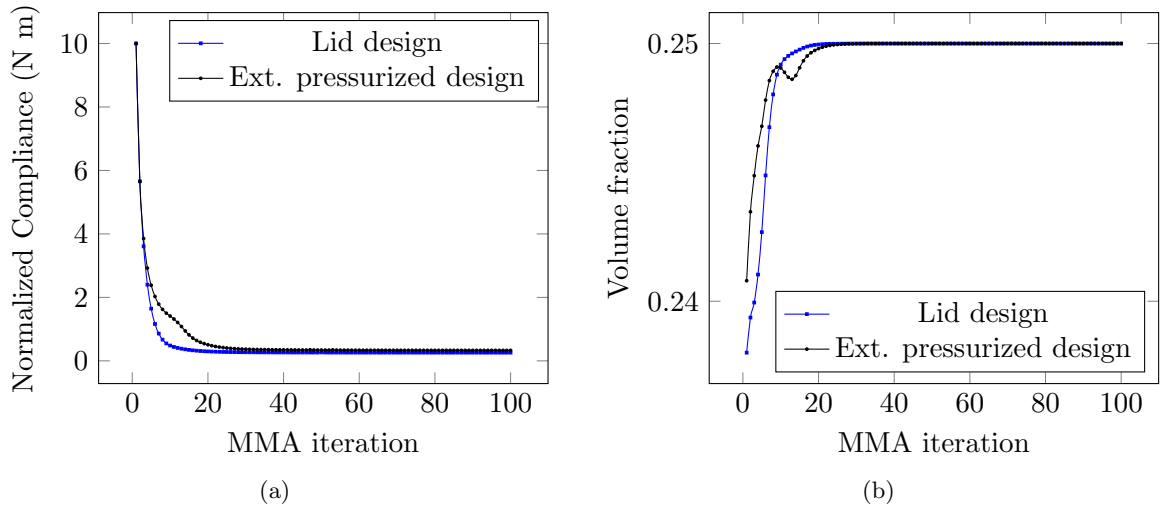

The optimized designs in different views for the lid and the externally pressurized  are depicted in Fig.~\ref{fig:Bathtubdesignresult} and Fig.~\ref{fig:Ex2result}, respectively. To plot the optimized results, an isosurface with the physical density value at 0.25 is used.  The exterior parts of the optimized design are shown in Fig.~\ref{fig:Bathtubview1} and Fig.~\ref{fig:Ex2view1}, respectively. In both cases, the optimizer has succeeded in reshaping the pressure-loaded surface into a configuration that is advantageous for the considered compliance objective. Material distributions for the optimized lid and externally pressurized loadbearing structures  with respect to different cross sections are displayed in Fig.~\ref{fig:EX_stiff_CSV}. One notices that the material densities in the cross sectional planes are close to 1.0, which indicates that the optimized designs converge towards 0-1 solutions (Fig.~\ref{fig:EX_stiff_CSV}). Near boundaries, intermediate densities are seen due to the density filtering. Nevertheless, the results allow for a clear design interpretation. The objectives and volume constraints convergence plots are illustrated via Fig.~\ref{fig:AllObjectiveplot} and Fig.~\ref{fig:AllVolumeplot}, respectively. It is found that the convergence plots are smooth and stable. The volume constraint remains active at the end of the optimization for each case and thus, the permitted volume is achieved.

\subsection{Pressure actuated compliant mechanisms}

Herein, three compliant mechanisms, e.g., inverter,  gripper and  magnifier are designed in 3D involving  design-dependent pressure loads using the multi-criteria objective, using the formulation given in Eq.~\ref{Eq:Optimizationequation}. These problems have two symmetry planes which are exploited herein and thus, only quarter of the design domain is optimized for each mechanism. 
\begin{figure}[!h]
	\begin{subfigure}[t]{0.19\textwidth}
		\centering
		\includegraphics[width=\linewidth]{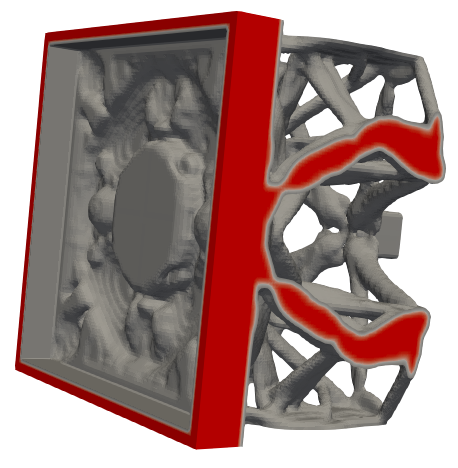}
		\caption{Arbitrary direction}
		\label{fig:CM1view1}
	\end{subfigure}
	\begin{subfigure}[t]{0.19\textwidth}
		\centering
		\includegraphics[width=\linewidth]{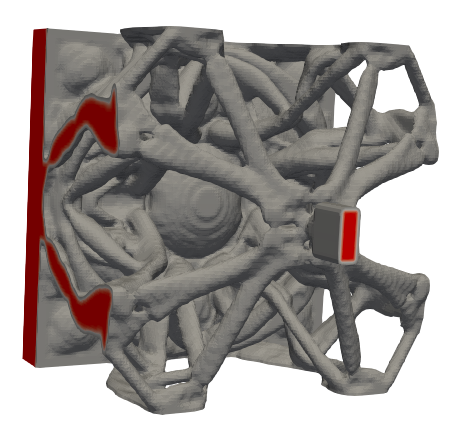}
		\caption{Arbitrary direction}
		\label{fig:CM1view1_1}
	\end{subfigure}
	\begin{subfigure}[t]{0.19\textwidth}
		\centering
		\includegraphics[width=\linewidth]{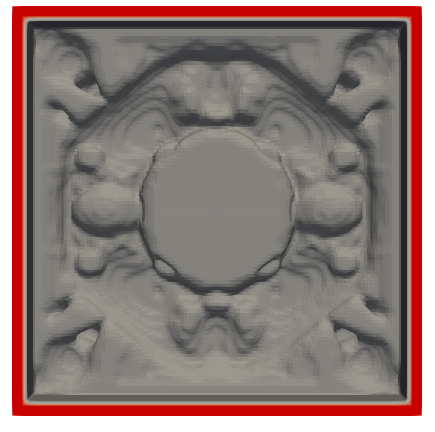}
		\caption{$+x-$direction}
		\label{fig:CM1view4}
	\end{subfigure}
	\begin{subfigure}[t]{0.19\textwidth}
		\centering
		\includegraphics[width=\linewidth]{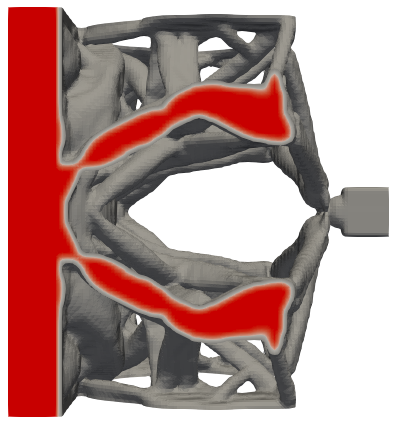}
		\caption{$+y-$direction}
		\label{fig:CM1view2}
	\end{subfigure}
	\begin{subfigure}[t]{0.19\textwidth}
		\centering
		\includegraphics[width=\linewidth]{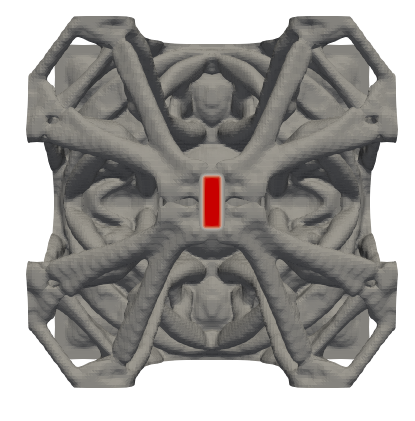}
		\caption{$-x-$direction}
		\label{fig:CM1view3}
	\end{subfigure}
	\caption{Optimized inverter design is shown in different view directions.}\label{fig:Inverter}
\end{figure}

\begin{figure}[!h]
	\begin{subfigure}[t]{0.19\textwidth}
		\centering
		\includegraphics[width=0.95\linewidth]{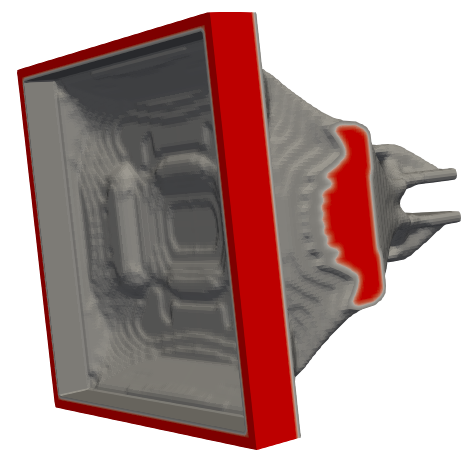}
		\caption{Arbitrary direction}
		\label{fig:CM2view1}
	\end{subfigure}
	\begin{subfigure}[t]{0.19\textwidth}
		\centering
		\includegraphics[width=0.95\linewidth]{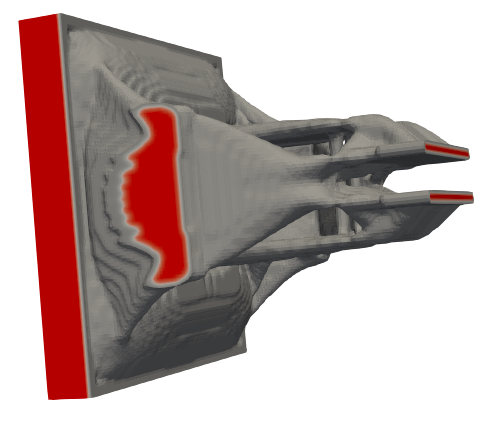}
		\caption{Arbitrary direction}
		\label{fig:CM2view1_1}
	\end{subfigure}
	\begin{subfigure}[t]{0.19\textwidth}
		\centering
		\includegraphics[width=0.90\linewidth]{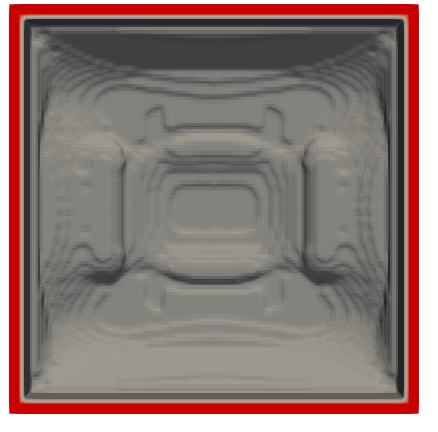}
		\caption{$+x-$direction}
		\label{fig:CM2view2}
	\end{subfigure}
	\begin{subfigure}[t]{0.19\textwidth}
		\centering
		\includegraphics[width=\linewidth]{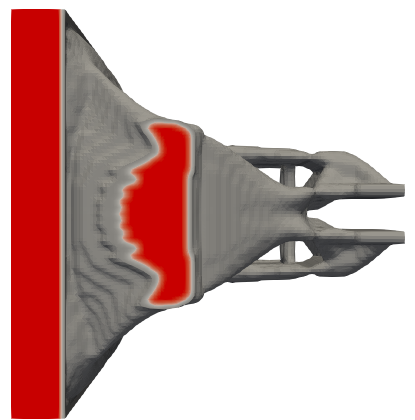}
		\caption{$+y-$direction}
		\label{fig:CM2view3}
	\end{subfigure}
	\begin{subfigure}[t]{0.19\textwidth}
		\centering
		\includegraphics[width=\linewidth]{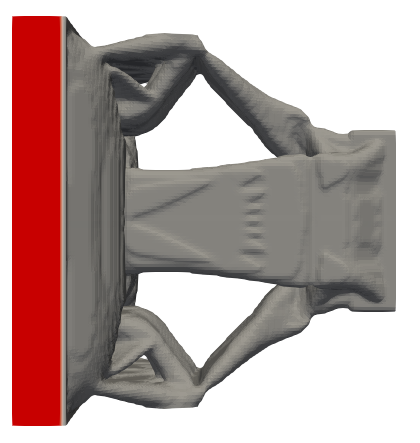}
		\caption{$-z-$direction}
		\label{fig:CM2view4}
	\end{subfigure}
	\caption{Optimized gripper design is shown in different view directions.}\label{fig:Gripper}
\end{figure}

\begin{figure}[!h]
	\begin{subfigure}[t]{0.19\textwidth}
		\centering
		\includegraphics[width=\linewidth]{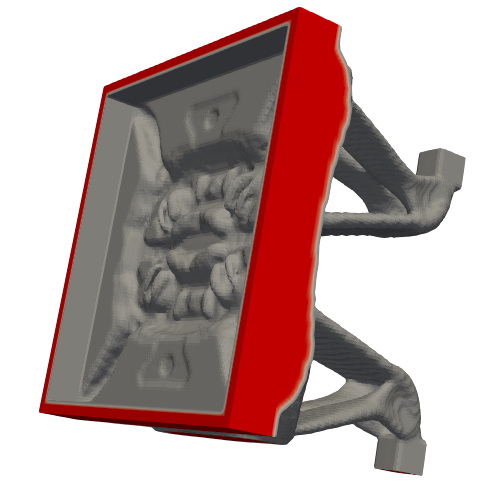}
		\caption{Arbitrary direction}
		\label{fig:CM3view1}
	\end{subfigure}
	\begin{subfigure}[t]{0.19\textwidth}
		\centering
		\includegraphics[width=\linewidth]{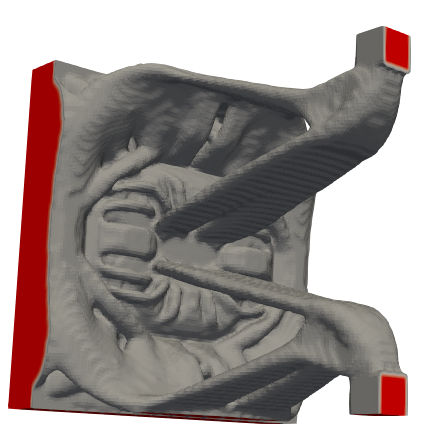}
		\caption{Arbitrary direction}
		\label{fig:CM3view1.1}
	\end{subfigure}
	\begin{subfigure}[t]{0.19\textwidth}
		\centering
		\includegraphics[width=\linewidth]{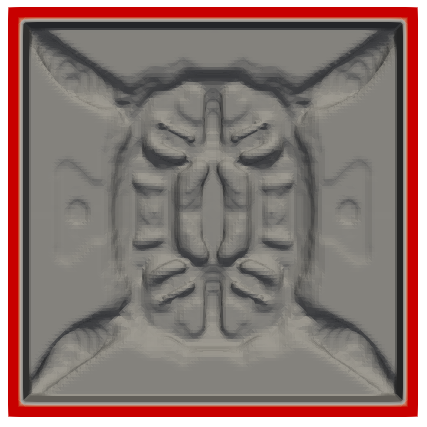}
		\caption{$+x-$direction}
		\label{fig:CM3view2}
	\end{subfigure}
	\begin{subfigure}[t]{0.19\textwidth}
		\centering
		\includegraphics[width=\linewidth]{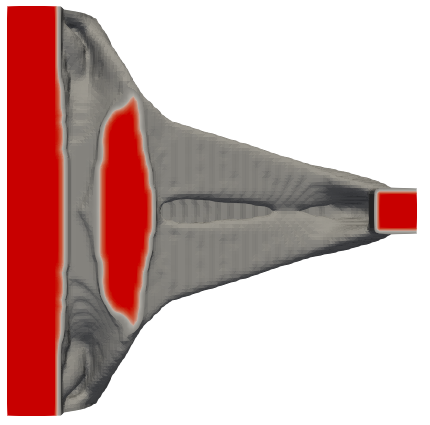}
		\caption{$+y$-direction}
		\label{fig:CM3view3}
	\end{subfigure}
	\begin{subfigure}[t]{0.19\textwidth}
		\centering
		\includegraphics[width=\linewidth]{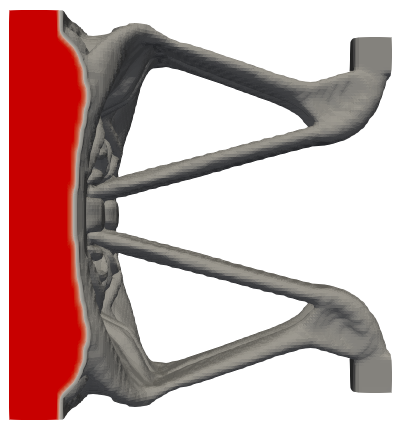}
		\caption{$-z-$direction}
		\label{fig:CM3view4}
	\end{subfigure}
	\caption{Optimized magnifier design is shown in different view directions.}\label{fig:Maginfier}
\end{figure}
Figure~\ref{fig:DesigndomainCM_inverter}, Fig.~\ref{fig:DesigndomainCM_gripper} and Fig.~\ref{fig:DesigndomainCM_magnifier} show the design specifications for one quarter mechanism designs.  Symmetry planes are also depicted. An inlet pressure load of 1.0 bar is applied from the left face of each mechanism design domain, whereas apart from symmetric faces other remaining faces experience  zero pressure load. Again as in the previous examples, instead of using a predetermined pressurized surface, the location and shape of the pressurized structural surface is subject to design optimization using the proposed formulation. Dimensions of each mechanism are set to ${0.2}\times{0.2}\times\SI{0.2}{\meter\cubed}$. We use  $120\times 60\times 60$ FEs to describe the considered quarter of each mechanism domain. The permitted volume fraction for each mechanism is set to 0.1. A rim of solid non-design region with size $\frac{L_x}{8}\times\frac{L_y}{15}\times\frac{L_z}{15}$ is considered around the pressure inlet area in each mechanism design, indicating its maximum size. To contain the applied pressure loading, a void non-design domain of maximum size $\frac{L_x}{10}\times\frac{14L_y}{15}\times\frac{14L_z}{15}$ is considered in front of the loading. The step parameters for the flow  and drainage coefficients are set to $\eta_k=0.1$ and $\eta_d= 0.2$, respectively \cite{kumar2020topology}. The scaling factor for the objective is set to $\mu=100$. A unit dummy load is applied along the desired  deformation direction of the mechanism to facilitate evaluation of the mutual strain-energy. For the quarter gripper design, a jaw (solid passive domain) of size $\frac{L_x}{8}\times\frac{L_y}{2}\times\frac{L_z}{20}$ is considered above a void non-design region with size $\frac{L_x}{8}\times\frac{L_y}{1}\times\frac{L_z}{10}$. Each node of the jaw is connected to springs representing the workpiece with a stiffness of $\SI{50}{\newton\per\meter}$. The desired gripping motion of the mechanism is in the $z-$direction. In case of the compliant inverter and magnifier mechanisms, the respective workpiece is represented via springs of stiffness $\SI{500}{\newton\per\meter}$. The desired motion for the inverter mechanism is in the negative $x-$direction, whereas for the magnifier an outward movement in the $y-$direction is sought.

\begin{figure}[!h]
	\begin{subfigure}[t]{0.32\textwidth}
		\centering
		\includegraphics[width=\linewidth]{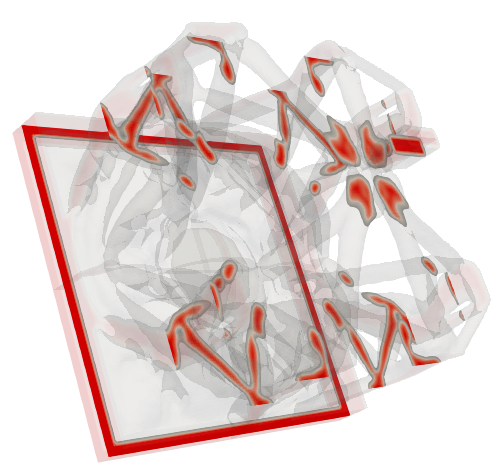}
		\caption{}
		\label{fig:CM1CSV}
	\end{subfigure}
	\begin{subfigure}[t]{0.32\textwidth}
		\centering
		\includegraphics[width=\linewidth]{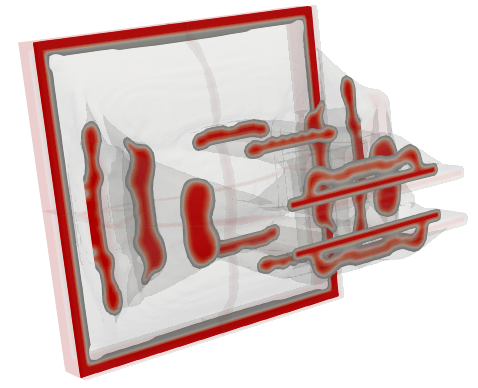}
		\caption{}
		\label{fig:CM2CSV}
	\end{subfigure}
	\begin{subfigure}[t]{0.32\textwidth}
		\centering
		\includegraphics[width=\linewidth]{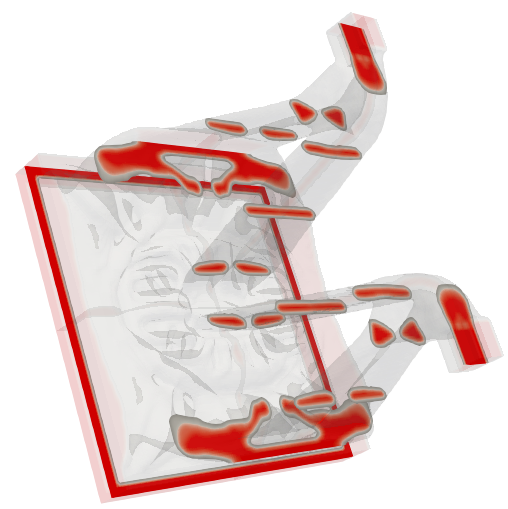}
		\caption{}
		\label{fig:CM3CSV}
	\end{subfigure}
	\caption{Material distributions of the optimized inverter, gripper and magnifier compliant mechanisms with respect to different cross sectional planes in arbitrary directions are  displayed in \subref{fig:CM1CSV}, \subref{fig:CM2CSV} and \subref{fig:CM3CSV}, respectively.}\label{fig:CMCSV}
\end{figure}

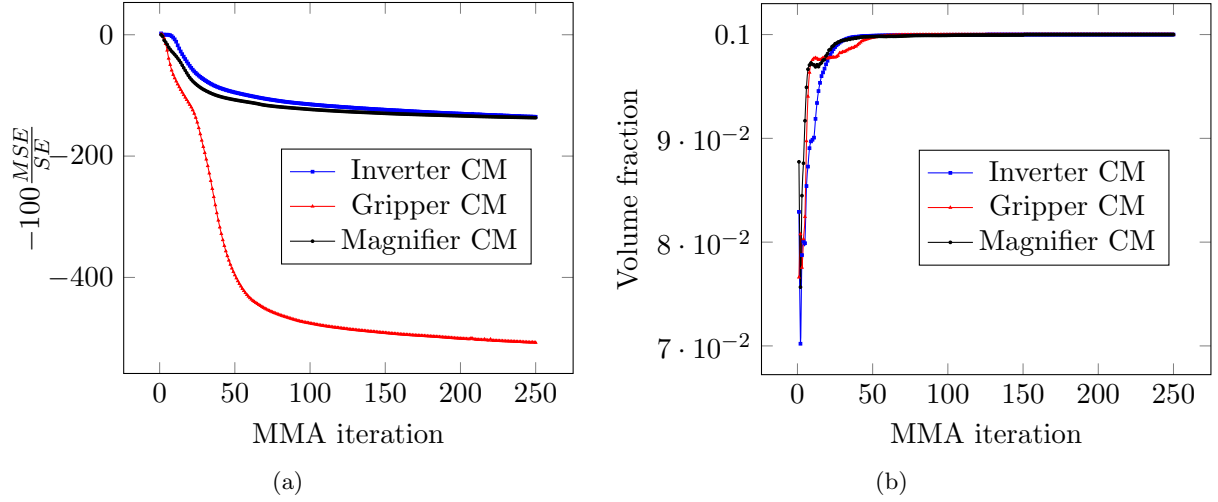
\begin{figure}[h!]
	\begin{subfigure}[t]{0.5\textwidth}
		\centering
		\begin{tikzpicture} 	
		\begin{axis}[
		width = 1\textwidth,
		xlabel=MMA iteration,
		ylabel= $-100\frac{MSE}{SE}$,
		legend style={at={(0.35,0.450)},anchor=west}]
		\pgfplotstableread{CM1Objective.txt}\mydata;
		\addplot[smooth,mark=square*,blue,,mark size=0.5pt]
		table {\mydata};
		\addlegendentry{Inverter CM}
		\pgfplotstableread{CM2Objective.txt}\mydata;
		\addplot[smooth,mark=triangle*,red,,mark size=0.5pt]
		table {\mydata};
		\addlegendentry{Gripper CM}
		\pgfplotstableread{CM3Objective.txt}\mydata;
		\addplot[smooth,,mark=*,black,mark size=0.5pt]
		table {\mydata};
		\addlegendentry{Magnifier CM}
		\end{axis}
		\end{tikzpicture}
		\caption{}
		\label{fig:AllObjectiveplotCM}
	\end{subfigure}
	\quad
	\begin{subfigure}[t]{0.5\textwidth}
		\centering
		\begin{tikzpicture}
		\pgfplotsset{compat = 1.3}
		\begin{axis}[
		width = 1\textwidth,
		xlabel=MMA iteration,
		ylabel= Volume fraction,
		legend style={at={(0.35,0.450)},anchor=west}]
		\pgfplotstableread{CM1Volume.txt}\mydata;
		\addplot[smooth,mark=square*,blue,,mark size=0.5pt]
		table {\mydata};
		\addlegendentry{Inverter CM}
		\pgfplotstableread{CM2Volume.txt}\mydata;
		\addplot[smooth,mark=triangle*,red,,mark size=0.5pt]
		table {\mydata};
		\addlegendentry{Gripper CM}
		\pgfplotstableread{CM3Volume.txt}\mydata;
		\addplot[smooth,mark=*,black,mark size=0.5pt]
		table {\mydata};
		\addlegendentry{Magnifier CM}
		\end{axis}
		\end{tikzpicture}
		\caption{}
		\label{fig:AllVolumeplotCM}
	\end{subfigure}
	\caption{Convergence objective and volume fraction plots for compliant mechanisms. (\subref{fig:AllObjectiveplotCM}) $-100\frac{MSE}{SE}$ history, and (\subref{fig:AllVolumeplotCM}) Volume fraction history.}
	\label{fig:CMConvergence}
\end{figure}

\begin{figure}[!h]
	\begin{subfigure}[t]{0.32\textwidth}
		\centering
		\includegraphics[width=\linewidth]{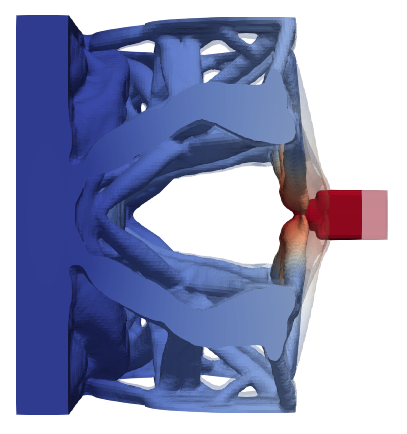}
		\caption{$+y-$direction}
		\label{fig:CM1deformed}
	\end{subfigure}
	\begin{subfigure}[t]{0.32\textwidth}
		\centering
		\includegraphics[width=\linewidth]{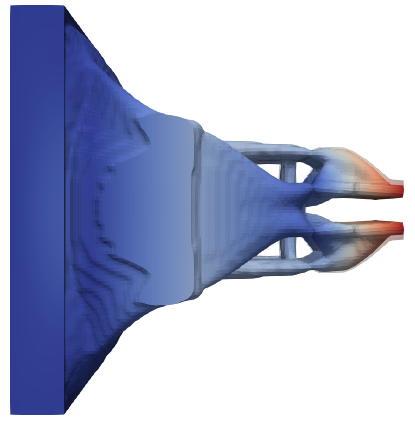}
		\caption{$+y-$direction}
		\label{fig:CM2deformed}
	\end{subfigure}
	\begin{subfigure}[t]{0.32\textwidth}
		\centering
		\includegraphics[width=\linewidth]{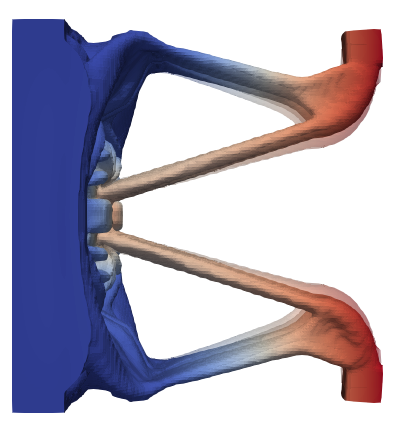}
		\caption{$-z-$direction}
		\label{fig:CM3deformed}
	\end{subfigure}
	\caption{Deformed profiles with respective view directions are shown with 500 times magnified displacements. The color scheme represents displacement field wherein red  and blue indicate maximum  and minimum displacements, respectively.}\label{fig:CMdeformed}
\end{figure}

The symmetric optimized results are transformed into respective final  full designs. Figure~\ref{fig:Inverter}, Fig.~\ref{fig:Gripper}, and Fig.~\ref{fig:Maginfier}, depict the 3D optimized designs in various views for the compliant inverter, gripper and magnifier mechanisms, respectively. The density value of the isosurface is displayed at 0.25.  While the TO process produces customized pressure-loaded membranes, that at the same time act as CMs themselves, the largest part of the design domain is filled with more or less traditional CM structures, that transmit and convert the pressure-induced deformations into the intended output deformations. In our experiments, we have not found cases where the majority of the design domain became filled with fluid. This is a clear difference from most pressure-loaded active structures as seen in, e.g., the field of soft robotics, where typically bellows-inspired designs are applied \cite{yap2016high}. It is noted, however, that the presented designs are based on linear structural analysis which is only valid for a limited deformation range. Note also that the pressurized membranes are not simply flat but contain corrugations and thicker and thinner regions. Similar to traditional compliant mechanisms, these geometries provide preferred deformation patterns that assist in the functioning of the mechanisms. The material distributions with respect to the different cross sections for the optimized inverter, gripper and magnifier mechanisms are illustrated in Fig.~\ref{fig:CM1CSV}, , which show that the structures have converged to clear solid-void designs, within the limits of the applied density filter. The convergence plots for the objectives and volume constraints are illustrated in Fig.~\ref{fig:CMConvergence}. One can notice the convergence history plots are smooth and stable. The volume constraint for each mechanism design is satisfied and active at the end of the optimization.

 Figure~\ref{fig:CMdeformed} displays the deformed profiles of the mechanisms. It is seen that in all cases the intended motion is produced. Note that because linear mechanical analysis is used, scaling of deformations is possible within a certain range. To reach deformations comparable to the design domain characteristic length, i.e., large deformation, one needs to consider nonlinear mechanics  within the topology optimization setting with high pressure loading, which is left for future research. This also requires configuration-dependent updating of the applied pressure loads, which could be achieved by solving Eqs.~\ref{Eq:Globalpressureequation} and \ref{Eq:GloablForcepressureconversion} on the deformed mesh. While the computational cost of these steps is small compared to the deformation analysis, the two problems become bidirectionally coupled and possibly a monolithic approach is preferred. Also sensitivity analysis of this coupled problem needs further study. 
 
\subsection{Computational cost}
As mentioned earlier, we employ the conjugate gradient algorithm with an incomplete Cholesky preconditioner in MATLAB to solve the 3D TO problems. Herein, we present the computational cost involved in the objective and sensitivity evaluation for the lid loadbearing structure and gripper mechanism with respect to the different mesh sizes, i.e., with different  number of design variables (NODVs) and degrees of freedom (DOFs). A 64-bit operating system machine with 16.0 GB RAM, Intel(R) Core(TM) i7-10700U CPU 2.90 GHz is used. For the objective evaluation (i) $\matr{A}\matr{p} = \matr{0}$, (ii) $\matr{F} = -\matr{D}\matr{p}$, (iii) $\mathbf{Ku = F}$ and (iv) $\mathbf{Kv = F_\mathrm{d}}$ are solved, and the corresponding computational cost is noted. Note that solution to (iv) is  needed only in case of CM problems. Displacement vectors $\mathbf{u}$ and $\mathbf{v}$ are further used in determining $\lambda_1$ and $\lambda_3$ (Eqs.~\ref{Eq:lagrangemultipliersstructure} and \ref{Eq:lagrangemultiplierCM}) for the sensitivity calculation. However, one needs to solve a system of equations to evaluate $\lambda_2$ and thus, sensitivity is evaluated using Eqs.~\ref{Eq:ComplianceSens} and \ref{Eq:CMSens}, and the corresponding computation time is recorded.

\begin{table}[h!]
	\begin{center}
			\caption{Computation time of objective and sensitivity evaluation  for lid loadbearing structure with different NODVs and DOFs.}\label{Table:computationalStructure}
			\begin{tabular}{c|c|c|c|c}
			\hline \hline
			\multirow{12}{*}{Lid design} & \multicolumn{1}{l|}{\multirow{2}{*}{NODVs}} & \multirow{2}{*}{DOFs}    & \multicolumn{2}{c}{Computation time (s)}                                               \\ \cline{4-5} 
			& \multicolumn{1}{l|}{}                       &                          & \multicolumn{1}{c|}{Objective evaluation} & \multicolumn{1}{c}{Sensitivity evaluation} \\ \cline{2-5} 
			& \multirow{2}{*}{16000}                      & \multirow{2}{*}{54243}   & \multicolumn{1}{c|}{\multirow{2}{*}{2.24}}    & \multicolumn{1}{c}{\multirow{2}{*}{0.060}}      \\
			&                                             &                          & \multicolumn{1}{c|}{}                     & \multicolumn{1}{c}{}                       \\ \cline{2-5} 
			& \multirow{2}{*}{54000}                      & \multirow{2}{*}{175863}  & \multirow{2}{*}{12.6}                         & \multirow{2}{*}{0.25}                           \\
			&                                             &                          &                                           &                                             \\ \cline{2-5} 
			& \multirow{2}{*}{128000}                     & \multirow{2}{*}{408483}  & \multirow{2}{*}{40.38}                         & \multirow{2}{*}{1.17}                           \\
			&                                             &                          &                                           &                                             \\ \cline{2-5} 
			& \multirow{2}{*}{250000}                     & \multirow{2}{*}{788103}  & \multirow{2}{*}{108.42}                         & \multirow{2}{*}{2.66}                           \\
			&                                             &                          &                                           &                                             \\ \cline{2-5} 
			& \multirow{2}{*}{432000}                     & \multirow{2}{*}{1350723} & \multirow{2}{*}{263.15}                         & \multirow{2}{*}{5.32}                           \\
			&                                             &                          &                                           &                                             \\ \hline \hline
	\end{tabular}
	\end{center}
\end{table}

\begin{table}[h!]
	\begin{center}
				\caption{Computation time of objective and sensitivity evaluation for the gripper mechanism with different NODVs and DOFs.}\label{Table:computationalCM}
			\begin{tabular}{c|c|c|c|c}
			\hline \hline
			\multirow{12}{*}{Gripper mechanism} & \multicolumn{1}{l|}{\multirow{2}{*}{NODVs}} & \multirow{2}{*}{DOFs}    & \multicolumn{2}{c}{Computation time (s)}                                               \\ \cline{4-5} 
			& \multicolumn{1}{l|}{}                       &                          & \multicolumn{1}{c|}{Objective evaluation} & \multicolumn{1}{c}{Sensitivity evaluation} \\ \cline{2-5} 
			& \multirow{2}{*}{16000}                      & \multirow{2}{*}{54243}   & \multicolumn{1}{c|}{\multirow{2}{*}{5.08}}    & \multicolumn{1}{c}{\multirow{2}{*}{0.33}}      \\
			&                                             &                          & \multicolumn{1}{c|}{}                     & \multicolumn{1}{c}{}                       \\ \cline{2-5} 
			& \multirow{2}{*}{54000}                      & \multirow{2}{*}{175863}  & \multirow{2}{*}{36.32}                         & \multirow{2}{*}{1.27}                           \\
			&                                             &                          &                                           &                                             \\ \cline{2-5} 
			& \multirow{2}{*}{128000}                     & \multirow{2}{*}{408483}  & \multirow{2}{*}{170.60}                         & \multirow{2}{*}{3.52}                           \\
			&                                             &                          &                                           &                                             \\ \cline{2-5} 
			& \multirow{2}{*}{250000}                     & \multirow{2}{*}{788103}  & \multirow{2}{*}{634.13}                         & \multirow{2}{*}{8.52}                           \\
			&                                             &                          &                                           &                                             \\ \cline{2-5} 
			& \multirow{2}{*}{432000}                     & \multirow{2}{*}{1350723} & \multirow{2}{*}{1769.15}                         & \multirow{2}{*}{14.03}                           \\
			&                                             &                          &                                           &                                             \\ \hline \hline
	\end{tabular}
	\end{center}
\end{table}

The computational expenses for the loadbearing structure and gripper CM are displayed in Table~\ref{Table:computationalStructure} and Table~\ref{Table:computationalCM}, respectively. One can notice that evaluation of the objective is more expensive than that of the sensitivity calculation, for the reasons discussed above. In addition, it can be noted that objective evaluation for the CM needs comparatively more time than that of loadbearing structures for the same number of NODVs and DOFs.\\

	\section{Closure}\label{Sec:Closure}
This paper presents a density-based topology optimization approach for designing design-dependent pressure-actuated (loaded) small deformation 3D compliant mechanisms and 3D loadbearing structures. The efficacy and versatility of the method in the 3D case is demonstrated  by designing various pressure-loaded 3D structures (lid and externally pressurized design) and pressure-actuated small deformation 3D compliant mechanisms (inverter, gripper and magnifier). For a loadbearing structure, compliance is minimized whereas a multi-criteria objective is employed for designing CMs.

The Darcy law in association with a drainage term is employed to convert the applied pressure loads into a design-dependent pressure field wherein the flow coefficient of an FE is related to its design variable using a smooth Heaviside function. It has been illustrated how the drainage term with the Darcy flux gives an appropriate pressure field for a 3D TO setting.  The presented approach provides a continuous pressure field which is converted into consistent nodal forces using a transformation matrix.  

The method finds pressure loading surfaces implicitly as topology optimization evolves and also, facilitates easy and computationally cheap evaluation of the load sensitivities using the adjoint-variable method. As pressure loading changes its location and magnitude, it is important to consider the load sensitivity terms while evaluating the objective sensitivity. For the presented numerical examples, it is noted that the objective evaluation is  computationally more expensive than sensitivity calculation.  The obtained 3D pressure-actuated mechanisms resemble a combination of a tailored pressurized membrane for load transfer, and a more conventional compliant mechanism design involving flexure hinges. It is suggested that different design solutions may emerge once larger deflections can be included. Extension of the approach with nonlinear continuum mechanics is therefore one of the prime directions for future work.

	\section*{Acknowledgments} 
	The authors would like to gratefully acknowledge Prof. Ole Sigmund for his suggestions and thanks  Prof. Krister Svanberg for providing MATLAB codes of the MMA optimizer.
	\numberwithin{equation}{section}
	\section*{Flow contrast}\label{Sec:Flowcontrast}
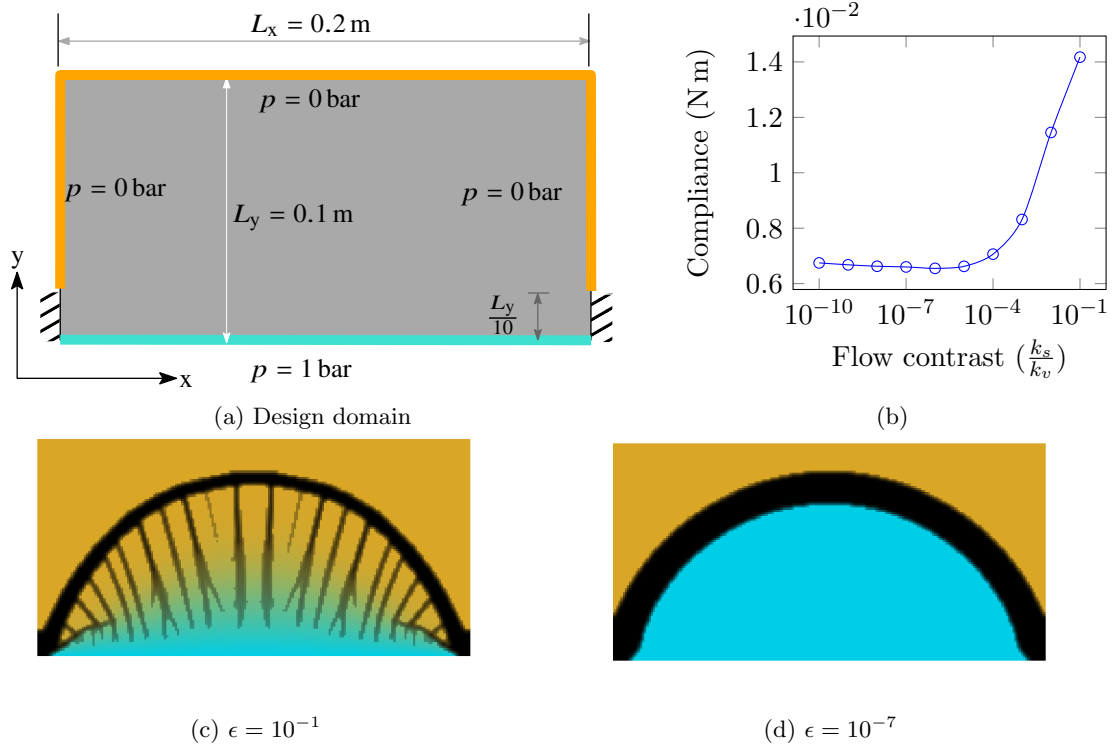
\begin{figure}[h!]
	\begin{subfigure}[t]{0.6\textwidth}
		\centering
		\includegraphics[scale = 1.25]{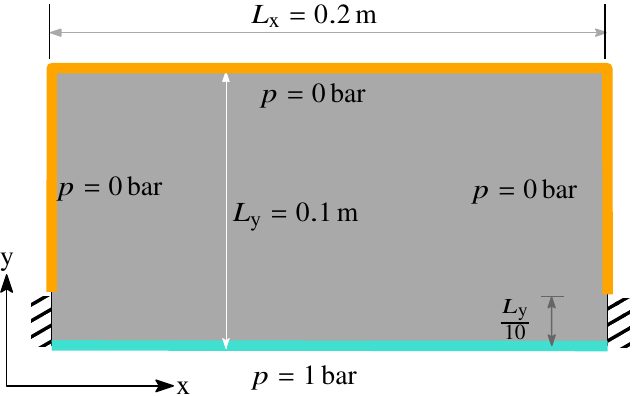}
		\caption{Design domain}
		\label{fig:2Darcproblemdesign}
	\end{subfigure}
	\begin{subfigure}[t]{0.38\textwidth}
	\centering
	\begin{tikzpicture}
	\centering
	\begin{semilogxaxis}[
	width = \textwidth,
	xlabel=Flow contrast ($\frac{k_s}{k_v}$),
	ylabel= Compliance ($\si{\newton\meter}$)]
	\addplot[smooth,mark=o,blue] plot coordinates {
		(1e-1,	0.014172302485361) (1e-2,0.011454737276392) (1e-3,	0.008315364784439)
		(1e-4,	0.007063706316265) (1e-5,	0.006623298729294) (1e-6,	0.006550838705038)
		(1e-7,	0.006602099271926) 	(1e-8,	0.006627030673618) (1e-9,	0.006679207275892)
		(1e-10,	0.006747771021997)
	};
	\end{semilogxaxis}
	\end{tikzpicture}
	\caption{}
	\label{fig:flowcontrast}
\end{subfigure}
	\begin{subfigure}[t]{0.5\textwidth}
		\centering
		\includegraphics[scale=0.5]{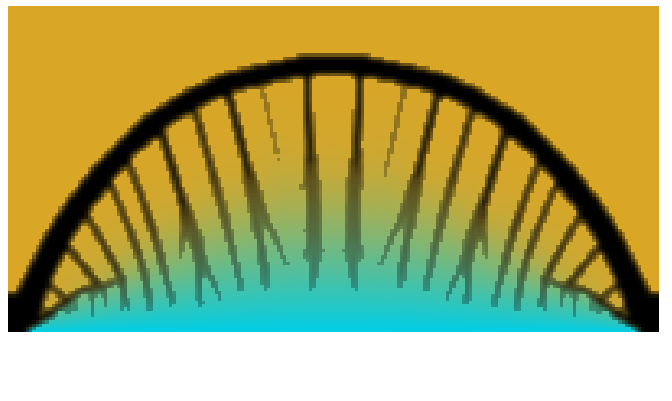}
		\caption{ $\epsilon=10^{-1}$}
		\label{fig:pressurefielde01}
	\end{subfigure}
	\begin{subfigure}[t]{0.5\textwidth}
		\centering
		\includegraphics[scale=0.5]{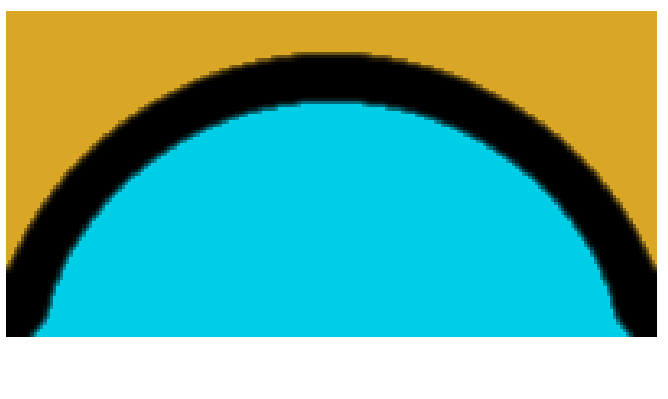}
		\caption{$\epsilon=10^{-7}$}
		\label{fig:pressurefielde07}
	\end{subfigure}
	\caption{(\subref{fig:2Darcproblemdesign}) Design domain for 2D internally pressurized arc. The optimized results with final pressure distribution using $\epsilon=0.1$ and  $\epsilon= 10^{-7}$ are shown in (\subref{fig:pressurefielde01}) and (\subref{fig:pressurefielde07}), respectively.}\label{fig:flowconstrast}
\end{figure}

Herein, an additional test problems is presented to illustrate the influence of flow contrast $\epsilon$ using an internally pressurized arc design  (Fig.~\ref{fig:2Darcproblemdesign}). We consider a 2D setting for the sake of simplicity and ease of result visualization, but the findings extend naturally to the 3D case. The design domain is described via $N_\text{ex} \times N_\text{ey} = 200\times100$ bi-linear rectangular FEs, where $N_\text{ex}$ and $N_\text{ey}$ represent the number of FEs in the $x-$ and $y-$directions, respectively. The filter radius and volume fraction are set to $2\times\min({\frac{L_x}{N_\text{ex}},\frac{L_y}{N_\text{ey}}})$, and 0.2, respectively. The maximum number of iterations for the optimization is set to 100. The design parameters mentioned in Table~\ref{Table:T1} are used. 

This optimization is performed for a range of  $\epsilon $ values. Fig.~\ref{fig:flowcontrast} depicts the convergence curve for the compliance objective minimization with the different flow contrasts. As examples, Figs.~\ref{fig:pressurefielde01} and \ref{fig:pressurefielde07} depict final solutions with respective pressure fields obtained using flow contrast $\epsilon=10^{-1}$ and $\epsilon=10^{-7}$, respectively. In Fig.~\ref{fig:pressurefielde01} it can be seen that also in void regions, a clear pressure gradient occurs. This is a direct result of the low flow contrast (Eq.~\ref{Eq:Globalpressureequation}). Since a pressure gradient leads to nodal force contributions (Eq.~\ref{Eq:GloablForcepressureconversion}), the optimization process creates semi-dense structures to increase the stiffness of the loaded regions, in order to minimize the total compliance. However, this is not a practical or realistic solution. These artifacts disappear with increased $\epsilon$.  Based on this study, we recommend that $\frac{K_\text{s}}{K_\text{v}}\in$ [$10^{-5},\,10^{-8}$]. In all other numerical examples in this paper, $\epsilon=10^{-7}$ has been used.

\bibliography{myreference}

\begin{thebibliography}{10}

\bibitem{sigmund2013topology}
O.~Sigmund and K.~Maute, ``Topology optimization approaches,'' {\em Structural
  and Multidisciplinary Optimization}, vol.~48, no.~6, pp.~1031--1055, 2013.

\bibitem{kenway2014multipoint}
G.~K. Kenway and J.~R. Martins, ``Multipoint high-fidelity aerostructural
  optimization of a transport aircraft configuration,'' {\em Journal of
  Aircraft}, vol.~51, no.~1, pp.~144--160, 2014.

\bibitem{rus2015design}
D.~Rus and M.~T. Tolley, ``Design, fabrication and control of soft robots,''
  {\em Nature}, vol.~521, no.~7553, pp.~467--475, 2015.

\bibitem{yang2016overview}
C.~Yang and F.~Huang, ``An overview of simulation-based hydrodynamic design of
  ship hull forms,'' {\em Journal of Hydrodynamics}, vol.~28, no.~6,
  pp.~947--960, 2016.

\bibitem{kumar2020topology}
P.~Kumar, J.~S. Frouws, and M.~Langelaar, ``Topology optimization of fluidic
  pressure-loaded structures and compliant mechanisms using the darcy method,''
  {\em Structural and Multidisciplinary Optimization}, pp.~1--19, 2020.

\bibitem{Hammer2000}
V.~B. Hammer and N.~Olhoff, ``{Topology optimization of continuum structures
  subjected to pressure loading},'' {\em Structural and Multidisciplinary
  Optimization}, vol.~19, no.~2, pp.~85--92, 2000.

\bibitem{du2004topologicalb}
J.~Du and N.~Olhoff, ``Topological optimization of continuum structures with
  design-dependent surface loading--part {II}: algorithm and examples for {3D}
  problems,'' {\em Structural and Multidisciplinary Optimization}, vol.~27,
  no.~3, pp.~166--177, 2004.

\bibitem{zhang2010topology}
H.~Zhang, S.-T. Liu, and X.~Zhang, ``Topology optimization of {3D} structures
  with design-dependent loads,'' {\em Acta Mechanica Sinica}, vol.~26, no.~5,
  pp.~767--775, 2010.

\bibitem{ananthasuresh1994methodical}
G.~K. Ananthasuresh, S.~Kota, and Y.~Gianchandani, ``A methodical approach to
  the design of compliant micromechanisms,'' in {\em Solid-state sensor and
  actuator workshop}, vol.~1994, pp.~189--192, SC: IEEE, 1994.

\bibitem{frecker1997topological}
M.~Frecker, G.~K. Ananthasuresh, S.~Nishiwaki, N.~Kikuchi, and S.~Kota,
  ``Topological synthesis of compliant mechanisms using multi-criteria
  optimization,'' {\em Journal of Mechanical design}, vol.~119, no.~2,
  pp.~238--245, 1997.

\bibitem{sigmund1997design}
O.~Sigmund, ``On the design of compliant mechanisms using topology
  optimization,'' {\em Journal of Structural Mechanics}, vol.~25, no.~4,
  pp.~493--524, 1997.

\bibitem{saxena2000optimal}
A.~Saxena and G.~K. Ananthasuresh, ``On an optimal property of compliant
  topologies,'' {\em Structural and multidisciplinary optimization}, vol.~19,
  no.~1, pp.~36--49, 2000.

\bibitem{saxena2001topology}
A.~Saxena and G.~K. Ananthasuresh, ``Topology synthesis of compliant mechanisms
  for nonlinear force-deflection and curved path specifications,'' {\em Journal
  of Mechanical Design}, vol.~123, no.~1, pp.~33--42, 2001.

\bibitem{pedersen2001topology}
C.~B. Pedersen, T.~Buhl, and O.~Sigmund, ``Topology synthesis of
  large-displacement compliant mechanisms,'' {\em International Journal for
  Numerical Methods in Engineering}, vol.~50, no.~12, pp.~2683--2705, 2001.

\bibitem{kumar2019compliant}
P.~Kumar, P.~Fanzio, L.~Sasso, and M.~Langelaar, ``Compliant fluidic control
  structures: Concept and synthesis approach,'' {\em Computers \& Structures},
  vol.~216, pp.~26--39, 2019.

\bibitem{kumar2019computational}
P.~Kumar, A.~Saxena, and R.~A. Sauer, ``Computational synthesis of large
  deformation compliant mechanisms undergoing self and mutual contact,'' {\em
  Journal of Mechanical Design}, vol.~141, no.~1, p.~012302, 2019.

\bibitem{kumar2020topologybio}
P.~Kumar, C.~Schmidleithner, N.~Larsen, and O.~Sigmund, ``Topology optimization
  and {3D} printing of large deformation compliant mechanisms for straining
  biological tissues,'' {\em Structural and Multidisciplinary Optimization},
  vol.~63, no.~3, pp.~1351--1366, 2021.

\bibitem{kumar2020topologyshape}
P.~Kumar, R.~A. Sauer, and A.~Saxena, ``On topology optimization of large
  deformation contact-aided shape morphing compliant mechanisms,'' {\em
  Mechanism and Machine Theory}, vol.~156, p.~104135, 2020.

\bibitem{du2004topologicala}
J.~Du and N.~Olhoff, ``Topological optimization of continuum structures with
  design-dependent surface loading--part {I}: new computational approach for
  {2D} problems,'' {\em Structural and Multidisciplinary Optimization},
  vol.~27, no.~3, pp.~151--165, 2004.

\bibitem{Fuchs2004}
M.~B. Fuchs and N.~N.~Y. Shemesh, ``{Density-based topological design of
  structures subjected to water pressure using a parametric loading surface},''
  {\em Structural and Multidisciplinary Optimization}, vol.~28, no.~1,
  pp.~11--19, 2004.

\bibitem{yang2005evolutionary}
X.~Yang, Y.~Xie, and G.~Steven, ``Evolutionary methods for topology
  optimisation of continuous structures with design dependent loads,'' {\em
  Computers \& structures}, vol.~83, no.~12-13, pp.~956--963, 2005.

\bibitem{Sigmund2007}
O.~Sigmund and P.~M. Clausen, ``{Topology optimization using a mixed
  formulation: An alternative way to solve pressure load problems},'' {\em
  Computer Methods in Applied Mechanics and Engineering}, vol.~196, no.~13-16,
  pp.~1874--1889, 2007.

\bibitem{Panganiban2010}
H.~Panganiban, G.~W. Jang, and T.~J. Chung, ``{Topology optimization of
  pressure-actuated compliant mechanisms},'' {\em Finite Elements in Analysis
  and Design}, vol.~46, no.~3, pp.~238--246, 2010.

\bibitem{zienkiewicz2005finite}
O.~C. Zienkiewicz and R.~L. Taylor, {\em The Finite Element Method for Solid
  and Structural Mechanics}.
\newblock Butterworth-heinemann, 2005.

\bibitem{bourdin2001filters}
B.~Bourdin, ``Filters in topology optimization,'' {\em International Journal
  for Numerical Methods in Engineering}, vol.~50, no.~9, pp.~2143--2158, 2001.

\bibitem{svanberg1987method}
K.~Svanberg, ``The method of moving asymptotes$-$a new method for structural
  optimization,'' {\em International Journal for Numerical Methods in
  Engineering}, vol.~24, no.~2, pp.~359--373, 1987.

\bibitem{yap2016high}
H.~K. Yap, H.~Y. Ng, and C.-H. Yeow, ``High-force soft printable pneumatics for
  soft robotic applications,'' {\em Soft Robotics}, vol.~3, no.~3,
  pp.~144--158, 2016.

\end{thebibliography}
\bibliographystyle{hieeetr}
\end{document}